\documentclass[10pt,aps,pra,twocolumn,superscriptaddress]{revtex4}

\usepackage{amsmath}
\usepackage{amsfonts}
\usepackage{amssymb}
\usepackage{multirow}
\usepackage{verbatim}
\usepackage{alltt}
\usepackage{moreverb}
\usepackage{graphicx,color,graphics}
\usepackage{hyperref}
\usepackage{setspace}
\usepackage{braket}
\usepackage{calligra}
\usepackage{bm}
\usepackage{titlesec}
\usepackage[normalem]{ulem}

\usepackage{amsmath,lipsum}    
\usepackage{graphicx}   
\usepackage{verbatim}   
\usepackage{color}      
\usepackage{subfigure}  
\usepackage{hyperref}   
\usepackage{setspace,enumerate,amssymb,amsthm,verbatim}
\usepackage{mathtools}
\usepackage[normalem]{ulem}
\usepackage{natbib}

\titleformat{\paragraph}
{\itshape\small}{\theparagraph}{1em}{}
\titlespacing{\paragraph}
{80pt plus 4pt minus 2pt}{0.8ex plus .1ex}{0.8ex plus .1ex}

\newtheorem{theorem}{Observation}

\providecommand{\ignore}[1]{}

\DeclareMathOperator{\Tr}{Tr}

\begin{document}
\singlespacing

\title{Entanglement verification with detection-efficiency mismatch}

\author{Yanbao Zhang}
 \affiliation{Institute for Quantum Computing, University of Waterloo, Waterloo, Ontario N2L 3G1, Canada}
 \affiliation{Department of Physics and Astronomy, University of Waterloo, Waterloo, Ontario N2L 3G1, Canada}

\author{Norbert L\" utkenhaus}
 \affiliation{Institute for Quantum Computing, University of Waterloo, Waterloo, Ontario N2L 3G1, Canada}
 \affiliation{Department of Physics and Astronomy, University of Waterloo, Waterloo, Ontario N2L 3G1, Canada}

\date{\today}
\begin{abstract}
The security analysis of quantum key distribution is difficult to perform when there 
is efficiency mismatch between various threshold detectors involved in an experimental setup.
Even the verification that the device actually performs in the quantum domain, referred to 
as the task of entanglement verification, is hard to perform. In this article we provide 
such an entanglement-verification method for characterized detection-efficiency mismatch. 
Our method does not rely on a cut-off of photon numbers in the optical signal. It can 
be applied independently of the degrees of freedom involved, thus covering, for example, 
efficiency mismatch in polarization and time-bin modes, but also in spatial modes.
The evaluation of typical experimental scenarios suggests that an increase of 
detection-efficiency mismatch will drive the performance of a given setup out of 
the quantum domain.
\end{abstract}

\maketitle

\section{Introduction}
The ability to verify effective entanglement in observed data
is a necessary condition for Alice and Bob to perform secure quantum 
key distribution (QKD)~\cite{Marcos2004}, demonstrate quantum 
teleportation~\cite{QuTele1993} and realize quantum repeaters~\cite{QuRepeater1998}. 
Many methods have been exploited for verifying entanglement. For example, 
one can apply the positive partial-transpose (PPT) criterion~\cite{Peres1996,Horodecki1996}, 
construct symmetric extensions of a quantum state~\cite{DPS2002}, or build 
expectation-values matrices (EVMs) from experimental observations and apply 
corresponding entanglement criteria~\cite{Vogel2005, Johannes2006, NPA2007, 
Haseler2008, Moroder2013}.  Also, one can verify entanglement by directly 
measuring special observables, e.g., Bell inequalities~\cite{Bell1964} or 
entanglement witnesses~\cite{Terhal2000,EntWit2000}. (See the review 
paper~\cite{Otfried2009} for a list and discussions of various methods.) 
For the typical QKD scenarios where Alice sends optical signals
to Bob, we can model the underlying quantum state as Alice holding a discrete 
finite-dimensional system while Bob receiving an infinite-dimensional optical 
mode. In this case, EVM-based verification methods are particularly useful.  
They have been well developed~\cite{Johannes2006, Haseler2008, Haseler2009, Haseler2010, 
Killoran2011, Moroder2013} and applied to real experiments~\cite{Killoran2012,Khan2013}.

Almost all previously known methods for verifying entanglement assume 
that various threshold detectors involved in an experimental setup are 
ideal with perfect efficiency~\footnote{The method using Bell 
inequalities is an exception. It can be applied to the general case where 
the detectors involved have arbitrary different efficiencies. Unfortunately, 
violations of Bell inequalities are usually not robust against transmission 
loss which is more severe than detection inefficiency in practice. As a 
consequence, they cannot verify entanglement in many QKD setups.}. This 
assumption can be justified when there is no efficiency mismatch between 
these threshold detectors. In this case, one cannot distinguish no-detection 
events due to detection inefficiency from those due to transmission loss. 
For simplicity of analysis, one can lump these two kinds of loss together 
as a new increased transmission loss followed by ideal threshold detectors with 
perfect efficiency (see Sect.~\ref{renormalization} for detailed discussions 
about this treatment). Then, one can verify entanglement and further prove 
the security of the corresponding QKD protocols.  

However, in practice it is hard to build two detectors that have exactly the 
same efficiency (for example, due to different samples of the fabrication
process). In the presence of efficiency mismatch, one cannot treat detection 
inefficiency in the same way as transmission loss, and so previously known methods 
cannot be applied for entanglement verification.

A detection-efficiency mismatch can also be induced by an adversary using 
the fact that a detector can respond to a photon differently depending on 
degrees of freedom (for example, spatial mode, frequency, or arriving time) 
rather than those employed to encode information. If an adversary can 
control these degrees of freedom such that the induced efficiency mismatch is 
large enough, powerful attacks on QKD systems exist, as demonstrated in 
Refs.~\cite{Zhao2008} and~\cite{Gerhardt2011}. In typical experiments the 
efficiency mismatch may not seem significant, but it still means 
that the device cannot be covered by an existing security proof.

In this paper we develop a general method to verify entanglement in the 
presence of detection-efficiency mismatch. The method works as long as the 
efficiency mismatch is characterized, even if the mismatch depends on degrees 
of freedom of a photon that are not employed to encode information.  
We carefully study an implementation of the BB84-QKD prepare-and-measure 
protocol~\cite{BB84} with polarization encoding, where we take account 
of the fact that Bob receives signals in the infinite-dimensional mode space 
with no limit on the number of photons contained in that space. Our method 
is expected to work for other QKD protocols. The full security proof of QKD 
protocols with efficiency mismatch is still an open problem and is not addressed 
in the current paper either, though some essential tools developed here 
will carry over to such a security proof. Note that Ref.~\cite{Fung2009} 
studied the security proof of the BB84-QKD protocol with efficiency mismatch, 
under the additional assumption that Bob's system is a photonic qubit. However, 
this assumption is hard to justify in actual implementations of QKD 
where threshold detectors are being used.

The remainder of the paper is laid out as follows: In Sect.~\ref{exp_con}, we 
describe an experimental setup for implementing the BB84-QKD protocol and the 
efficiency-mismatch models considered. In Sect.~\ref{method}, we outline 
how our method works and explain details on how to apply the method to the 
particular experimental setup considered. In Sect.~\ref{results}, we simulate 
experimental results according to a toy channel connecting Alice and Bob just 
for illustration purposes. For this toy channel, we present the bound on the 
efficiency mismatch in order for Alice and Bob to verify entanglement based 
only on their observations. Finally we conclude the paper in Sect.~\ref{conclusion}.

\section{Experimental configuration} 
\label{exp_con}
For simplicity, in this paper we consider an experimental implementation of the 
BB84-QKD prepare-and-measure protocol~\cite{BB84} with polarization encoding. 
In each run of the protocol, Alice prepares an optical signal where all the 
photons have the same polarization, randomly selected from the horizontal ($H$), 
vertical ($V$), diagonal ($D$), or anti-diagonal ($A$) polarizations. Then, Alice 
sends the optical signal to Bob, and Bob randomly selects to measure it either in 
the horizontal/vertical ($H/V$) basis or the diagonal/anti-diagonal ($D/A$) basis.  
After many runs of the protocol, if the final measurement results satisfy some 
conditions (for example, the quantum bit error rate is low enough), Alice and Bob 
can distill secret keys via some classical post-processing procedure. For the present 
study of entanglement verification, we only need to consider the quantum phase of 
the protocol, i.e., the above prepare-and-measure step.

Obviously, in the above implementation there is no entangled state physically shared 
between Alice and Bob. However, there is another equivalent description, i.e., the 
source-replacement description~\cite{Marcos2004, BBM1992}, of the prepare-and-measure 
step in a general QKD protocol: In this thought setup, first Alice prepares an entangled state
\begin{equation}
\Ket{\Phi}_{AA'}=\sum_{s=1}^{S} \sqrt{p_s} \Ket{s}_A \Ket{\phi_s}_{A'}, 
\label{eq:ent_AB}
\end{equation}
where $\{\Ket{s}_A\}$ is a set of orthogonal states and $p_s$ is the probability 
of preparing the signal state $\Ket{\phi_s}$, $s=1,2,...,S$. Second, Alice measures 
the system $A$ with the projective positive-operator valued measure (POVM) 
$\{\Ket{s}\Bra{s}, s=1,2,..., S\}$, and distributes the corresponding signal state 
$\Ket{\phi_s}$ to Bob. After the action of the channel (or Eve) on system $A'$, 
Bob receives a system $B$ on which Bob performs a measurement. There is no way for 
Eve or any other party outside of Alice's lab to tell which description, either the 
prepare-and-measure or source-replacement description, is implemented at Alice's side. 
In the source-replacement description entanglement between Alice's system $A$ and 
Bob's system $B$ (before their respective measurement) is required; otherwise, 
intercept-resend attacks on the QKD system exist. In this sense, we say that 
effective entanglement is a necessary condition for secure QKD~\cite{Marcos2004}.

\begin{figure}
   \includegraphics[scale=0.58,viewport=5.5cm 14.5cm 16cm 20.5cm]{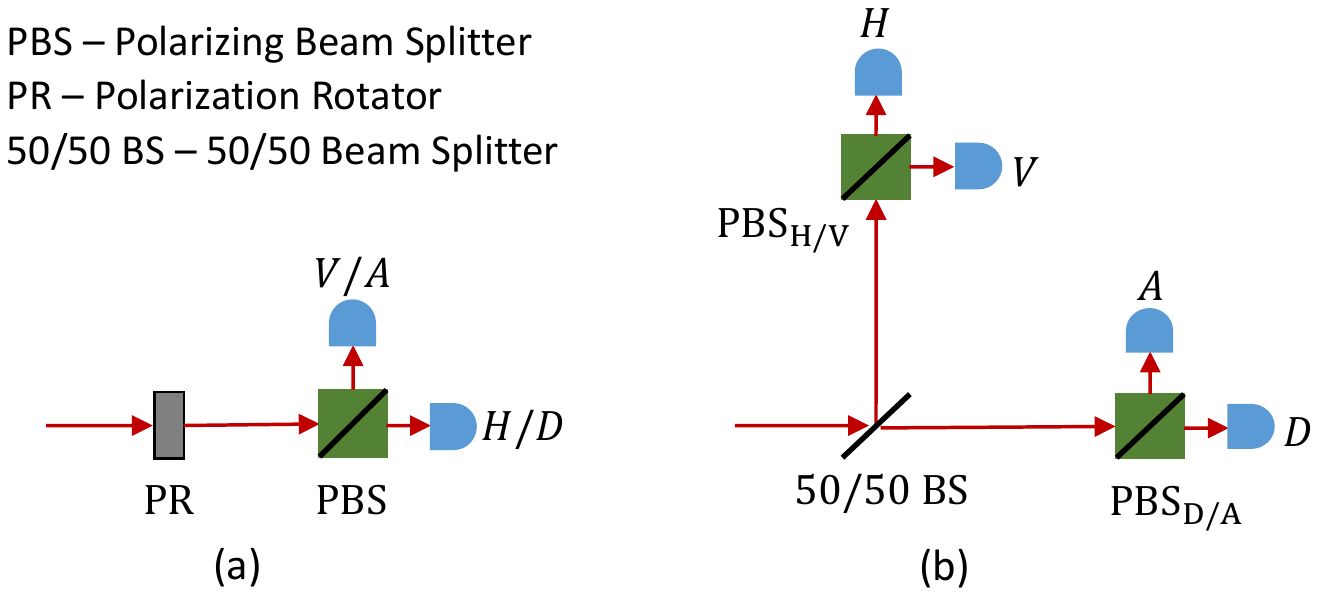} 
   \caption{Schematic of the measurement device: (a) is the active-detection
	 scheme where a polarization rotator is used to select a measurement basis,  
	 and (b) is the passive-detection scheme where a 50/50 beam splitter 
	 is used to select a measurement basis. Under each basis, 
	 a polarizing beam splitter and two threshold detectors (which 
	 cannot distinguish the number of incoming photons) are used to 
	 measure the polarization state of an incoming optical signal. Each 
	 detector is labelled by the corresponding measurement outcome.} 
   \label{setup} 
\end{figure}

To measure the polarization state of the incoming optical signal, Bob can 
employ either the active- or passive-detection scheme, as described in Fig.~\ref{setup}. 
The detectors involved in each detection scheme are threshold detectors which 
cannot distinguish the number of incoming photons. So, each detector has only 
two outcomes, click or no click. However, we do not restrict the number of 
photons arriving at each detector, due to the following two considerations: 
First, in practice information is usually encoded in coherent states which 
actually do have multi-photon components; second, the optical signal 
prepared by Alice can be intercepted by Eve and replaced by another stronger 
signal during the transmission from Alice to Bob.

In practice, the efficiencies of each detector are not exactly the 
same. For the active-detection scheme as shown in Fig.~\ref{setup}(a), the 
detection efficiency is denoted by $\eta_{H/D}$ if the measurement outcome 
is $H$ or $D$, and the efficiency is denoted by $\eta_{V/A}$ if the outcome 
is $V$ or $A$.  Similarly, for the passive-detection scheme as shown  
in Fig.~\ref{setup}(b), there are four detectors corresponding to the four  
measurement outcomes $H$, $V$, $D$ and $A$. Denote their respective efficiencies 
by $\eta_H$, $\eta_V$, $\eta_D$ and $\eta_A$. We will study entanglement  
verification in the presence of efficiency mismatch between these detectors.   
We call this the spatial-mode-independent mismatch model, in contrast to the 
following mismatch model where the mismatch depends additionally on the 
spatial modes.


The detection efficiency might not only be different between different 
detectors of the same build, but there might be also different coupling 
efficiencies of the detectors to the observed light. This has been 
demonstrated in recent works~\cite{Shihan2015} and~\cite{Rau2014}, where 
it has been shown that the coupling efficiency of each detector can be 
tuned to some degree independently by manipulating the spatial modes 
of the incoming optical signals.
As this kind of spatial-mode-dependent efficiency mismatch is 
quite relevant especially to implementations of free-space QKD, we would 
like to study the effect of this mismatch model on entanglement verification.   
In principle, the number of spatial modes of incoming optical signals can be 
arbitrary, even infinite.  Intuitively, when the number of spatial modes is 
equal to or larger than the number of detectors in the measurement device, it 
might become possible for Eve to completely control Bob through 
the efficiency mismatch. For example, if each detector responds to an optical 
signal only in a particular spatial mode and different detectors respond to 
different spatial modes, then Eve can completely effectively switch on 
and off each detector by sending optical signals in these particular spatial 
modes. To illustrate the effect of spatial-mode-dependent efficiency mismatch 
on entanglement verification, we consider the case where the number of spatial 
modes is equal to the number of detectors. For illustration purposes, we 
also constrain the mismatch model so that mismatched efficiencies have a 
permutation symmetry over spatial modes. In particular, the mismatch models 
considered in the active- and passive-detection schemes are shown as in 
Tables~\ref{active_mismatch} and \ref{passive_mismatch}, respectively. According 
to the discussion in Sect.~\ref{renormalization}, we can renormalize detection 
efficiencies and treat the common loss in detectors as a part of transmission loss.
Hence, we set the maximum detection efficiency in Tables~\ref{active_mismatch} 
or~\ref{passive_mismatch} to be $1$. We would like to stress that we consider 
these mismatch models just for simplicity and ease of graphical illustrations: 
The method detailed in the next section works for general mismatch models.

\begin{table}
\caption{Spatial-mode-dependent mismatch model in the active-detection scheme, 
where $0\leq \eta \leq 1$. Different columns are for different detectors labelled 
and shown in Fig.~\ref{setup}(a). Different rows are mismatched efficiencies for 
different spatial modes.}
\begin{center} 
\begin{tabular}  {|c |c|c|}

 \hline
 & Det. `$H/D$' & Det. `$V/A$'   \\ 
  \hline
 Mode 1 & 1 & $\eta$  \\
  \hline
 Mode 2 & $\eta$ & 1   \\
  \hline
\end{tabular}
\newline \\
\end{center}
\label{active_mismatch} 
\end{table}

\begin{table}
\caption{Spatial-mode-dependent mismatch model in the passive-detection scheme, 
where $0\leq \eta \leq 1$. Different columns are for different detectors labelled 
and shown in Fig.~\ref{setup}(b). Different rows are mismatched efficiencies for 
different spatial modes.}
\begin{center}
\begin{tabular}  {|c |c |c| c| c|}
 \hline 
 & Det. `$H$' & Det. `$V$' & Det. `$D$' & Det. `$A$' \\
  \hline
Mode 1 & 1 & $\eta$ & $\eta$ & $\eta$  \\
  \hline
 Mode 2 & $\eta$ & 1  & $\eta$ & $\eta$ \\
  \hline
 Mode 3 & $\eta$ & $\eta$ & 1 & $\eta$ \\
  \hline
 Mode 4 & $\eta$ & $\eta$ & $\eta$ & 1 \\
  \hline
\end{tabular}
\newline \\
\end{center}
\label{passive_mismatch} 
\end{table}

\section{Our method} 
\label{method}

The main idea behind our method is to construct an expectation-values 
matrix (EVM)~\cite{Johannes2006, Haseler2008,Haseler2009, Haseler2010,
Killoran2011, Moroder2013} using a finite number of actual measurement 
operators which contain efficiency-mismatch information. Let us discuss 
the construction and general properties of an EVM before moving on to 
our particular case. 

Suppose that the joint system of Alice and Bob is described by a state $\rho_{AB}$,
and that there are two sets of operators $\{\hat A_i\}$ and $\{\hat B_j\}$ 
acting on Alice's and Bob's subsystems, respectively.  Then, the entries of 
an EVM $\chi(\rho_{AB})$ are defined~\footnote{Strictly speaking, 
Eq.~\eqref{general_evm} defines only EVMs with a bipartite structure, which 
will be used for verifying entanglement shared between Alice and Bob.} as
\begin{equation}
[\chi(\rho_{AB})]_{ij,kl}=\text{Tr}(\rho_{AB} \hat A_i^{\dagger}\hat A_k \otimes \hat B_j^{\dagger}\hat B_l). 
\label{general_evm}
\end{equation}
By the definition, several properties are satisfied by an EVM: First,
an EVM is Hermitian and positive-semidefinite. Second,
if both the sets of operators $\{\hat A_i\}$ and $\{\hat B_j\}$ are finite, 
the dimension of the EVM $\chi(\rho_{AB})$ is finite even though the state 
$\rho_{AB}$ is infinite-dimensional. Third, entries of an EVM can be 
expectation values of observables, if the corresponding measurement 
operators are included in the set 
$\{ \hat A_i^{\dagger}\hat A_k \otimes \hat B_j^{\dagger}\hat B_l \}$.
Hence, an EVM will be designed as an object into which we can enter 
all experimental observations (i.e., the probabilities of Alice's 
and Bob's joint measurement outcomes), but there may be undetermined entries. Still, 
we can study various properties, such as entanglement, of the underlying state. 
Fourth, the linear relationships between various operators 
$\hat A_i^{\dagger}\hat A_k \otimes \hat B_j^{\dagger}\hat B_l$ restrict 
the entries $\chi_{ij,kl}$. (Here and later, we use $\chi$ and $\chi_{ij,kl}$ 
as short notations of $\chi(\rho_{AB})$ and $[\chi(\rho_{AB})]_{ij,kl}$ if 
there is no confusion in the context.) For example, if the operators satisfy 
\begin{equation}
\sum_{ijkl}C_{ij,kl} \hat A_i^{\dagger}\hat A_k \otimes \hat B_j^{\dagger}\hat B_l\geq 0,
\label{op_con}
\end{equation}
then for any density matrix $\rho_{AB}$ the entries of the corresponding EVM satisfy
\begin{equation}
\sum_{ijkl}C_{ij,kl} \chi_{ij,kl}\geq 0, 
\label{evm_con}
\end{equation}
where the coefficients $C_{ij,kl}$ are complex numbers. Eq.~\eqref{evm_con}
can be proved using the positivity of the whole operator in the left-hand side 
of Eq.~\eqref{op_con} and the definition of an EVM. We will exploit all the above 
properties to reduce the number of free parameters in the constructed EVM 
and so reduce the complexity of the entanglement-verification problem.

Furthermore, to verify entanglement we need the following observation~\cite{Haseler2008,Moroder2013}: 
\begin{theorem}
\label{evm_ppt}
If the state $\rho_{AB}$ is separable, then $\chi(\rho_{AB})$ has a separable 
structure and so satisfies the PPT criterion. 
\end{theorem}
This observation follows from the definition of an EVM and the PPT criterion~\cite{Peres1996,Horodecki1996} 
satisfied by all separable (and even un-normalized) states.
Note that an EVM is a un-normalized positive-semidefinite matrix.  
Thus, one can prove that the underlying state $\rho_{AB}$ must be entangled
by showing that the constructed EVM $\chi$ \emph{cannot} simultaneously satisfy the 
following two constraints: First, the entries $\chi_{ij,kl}$ are consistent with 
experimental observations and also with operator relationships; second, 
$\chi\geq 0$ and $\chi^{\Gamma_A}\geq 0$ where $\Gamma_A$ is the partial-transpose 
operation on Alice's system. Hence, entanglement verification can be 
formulated as a semidefinite programming (SDP) problem which can be solved 
efficiently (see Sect.~\ref{sdp} for details).

To construct an EVM useful for our situation, we need to choose 
appropriate sets of operators $\{\hat A_i\}$ and $\{\hat B_j\}$. 
In the following subsections, we will discuss in detail the set of operators 
that we consider in the case of efficiency mismatch, and also lay out several tricks 
that we can exploit to achieve our goal.

\subsection{Operators exploited for the construction of EVMs}
\label{evm_operators}

Let us consider Alice's side first. Recall that in the source-replacement 
description of a QKD protocol Alice first prepares the entangled state in 
Eq.~\eqref{eq:ent_AB} between systems $A$ and $A'$. Subsequently, 
Alice measures the system $A$ and sends the corresponding signal state 
encoded in system $A'$ to Bob. In the above process, the system $A$ 
remains at Alice. As a result, the reduced density matrix $\rho_A$ of 
system $A$ does not change even if the signal states change during 
the transmission from Alice to Bob. Also, Alice has complete knowledge of 
$\rho_A$, since the state in Eq.~\eqref{eq:ent_AB} is prepared and known 
by herself. Actually the overlap structure of signal states $\{\Ket{\phi_s}, s=1,2,...,S\}$ 
and the probabilities of preparing different signal states $\{p_s, s=1,2,...,S\}$ 
determine the reduced state $\rho_A$. The rank of the density matrix 
$\rho_A$ can be less than $S$, if the signal states prepared by Alice are 
linearly dependent. For example, in the ideal BB84-QKD protocol where 
information is encoded in the polarization of a single photon, the 
entangled state prepared by Alice is 
\begin{align}
\Ket{\Psi}_{AA'}=&\frac{1}{2}(\Ket{1}_A\Ket{H}_{A'}+\Ket{2}_A\Ket{V}_{A'} \notag \\
&+\Ket{3}_A\Ket{D}_{A'}+\Ket{4}_A\Ket{A}_{A'}),
\label{eq:ent_BB84}
\end{align}
where $\Ket{H},\Ket{V},\Ket{D}$, and $\Ket{A}$ are single-photon states
with horizontal, vertical, diagonal, and anti-diagonal polarizations, respectively. 
Although the reduced density matrix $\rho_A$ is of dimension $4\times 4$, it is easy 
to check that $\rho_A$ lives in a 2-dimensional subspace and that Alice's measurement 
$\{\Ket{1}\Bra{1}, \Ket{2}\Bra{2}, \Ket{3}\Bra{3}, \Ket{4}\Bra{4}\}$ also has a 
representation in the same 2-dimensional subspace. 

To take advantage of the complete knowledge of Alice's state $\rho_A$
and her measurement, we set the operators at Alice's side to be 
$\hat A_i=\Ket{\phi}\Bra{i}$ with $i=1,2,...,n$, where the pure states 
$\{\Ket{i}, i=1,2,...,n\}$ form a basis for the support of the density matrix 
$\rho_A$ and $\Ket{\phi}$ is an arbitrary pure state of Alice's system. By choosing 
these operators, we can make sure that Alice's state and observations (i.e., the 
probabilities of Alice's measurement results) all are included in the constructed 
EVM (if the set of operators considered at Bob's side contains the identity 
operator, which is usually a good choice). 

Now, let us proceed to Bob's side. Depending on which detection scheme in 
Fig.~\ref{setup} is used and which mismatch model is considered, the 
set of operators exploited is different. In the following subsection, 
we will discuss the operators exploited in the active-detection scheme 
with one spatial mode (i.e., the spatial-mode-independent scenario). 
The operators exploited in the other cases are postponed to 
Appendix~\ref{sect:operators} due to their complexities.

\subsubsection{A basic construction of EVMs}
\label{active_operators_1}

In the active-detection scheme as shown in Fig.~\ref{setup}(a), the 
four possible events in a measurement basis are click at only one of 
the two detectors (single click), clicks at both detectors (double 
click), and no click at neither of the two detectors. Suppose that 
the two detectors in Fig.~\ref{setup}(a) have efficiencies $\eta_{H/D}$ 
and $\eta_{V/A}$, respectively. Then, the POVM elements for both the 
$H/V$ and $D/A$ measurement choices can be written down explicitly. 
For example, the POVM elements for the measurement in the $H/V$ basis are 
\begin{align}
M_H=&\sum_{n_H=1}^{\infty}\sum_{n_V=0}^{\infty}\left(1-(1-\eta_{H/D})^{n_H}\right)(1-\eta_{V/A})^{n_V} \notag \\ 
&\Ket{n_H,n_V}\Bra{n_H,n_V}, \notag \\
M_V=&\sum_{n_H=0}^{\infty}\sum_{n_V=1}^{\infty}(1-\eta_{H/D})^{n_H}\left(1-(1-\eta_{V/A})^{n_V}\right) \notag \\
&\Ket{n_H,n_V}\Bra{n_H,n_V}, \notag \\
M_{HV}=&\sum_{n_H=1}^{\infty}\sum_{n_V=1}^{\infty}\left(1-(1-\eta_{H/D})^{n_H}\right)\left(1-(1-\eta_{V/A})^{n_V}\right)\notag \\
&\Ket{n_H,n_V}\Bra{n_H,n_V}, \text{ and} \notag\\
M_{\emptyset}^{+}=& \sum_{n_H=0}^{\infty}\sum_{n_V=0}^{\infty}(1-\eta_{H/D})^{n_H}(1-\eta_{V/A})^{n_V} \notag \\
&\Ket{n_H,n_V}\Bra{n_H,n_V}.
\label{eq:active_HV_POVMs}
\end{align}
Here, the subscripts of the POVM elements indicate the corresponding 
click events, the notation `$\emptyset$' means no click, the superscript 
`$+$' denotes the $H/V$ measurement basis, and $\Ket{n_H,n_V}$ is a photon-number 
basis state containing $n_H$ horizontally polarized photons and $n_V$ vertically polarized photons. 
See Appendix~\ref{sect:povm_with_mismatch} for the derivation of the 
POVM elements in Eq.~\eqref{eq:active_HV_POVMs}.  

The POVM elements in Eq.~\eqref{eq:active_HV_POVMs} satisfy two properties. 
First, it is obvious to see that these POVM elements are diagonal in the 
photon-number basis $\{\Ket{n_H,n_V}, n_H, n_V=0,1,2,...\}$. Hence, any two 
of them commute. Second, because $\eta_{H/D}$ and $\eta_{V/A}$ are between 0 
and 1, so are all coefficients of the terms $\Ket{n_H,n_V}\Bra{n_H,n_V}$ 
in Eq.~\eqref{eq:active_HV_POVMs}. Hence, we get the following relationships:
\begin{equation}
M_i\geq M_iM_j\geq 0, \label{eq:active_HV_rela}
\end{equation}
where $i,j=H, V$, or $HV$. Here, we write down $A \geq B$ when $(A-B)$ is a 
positive-semidefinite matrix. Using these two properties, we can restrict 
the entries of the constructed EVM if the measurement POVM elements in 
Eq.~\eqref{eq:active_HV_POVMs} are exploited. 

The above two properties are also satisfied by the POVM elements for the 
measurement in the $D/A$ basis. These POVM elements have the same expressions 
as those in Eq.~\eqref{eq:active_HV_POVMs} with the replacement of the 
subscripts $H$ and $V$ by $D$ and $A$, respectively. For example, the POVM 
element for the single-click event with diagonal polarization is 
\begin{align}
M_D=&\sum_{n_D=1}^{\infty}\sum_{n_A=0}^{\infty}\left(1-(1-\eta_{H/D})^{n_D}\right)(1-\eta_{V/A})^{n_A} \notag \\ 
&\Ket{n_D,n_A}\Bra{n_D,n_A}, \label{eq:active_DA_POVMs}
\end{align}
where the basis state $\Ket{n_D,n_A}$ contains $n_D$ diagonally polarized photons and $n_A$ 
anti-diagonally polarized photons. The expressions of the other three POVM elements $M_A, M_{DA}$
and $M_{\emptyset}^{\times}$, where the superscript `$\times$' indicates the $D/A$ basis, 
can be found in Appendix~\ref{sect:povm_with_mismatch}.  

The expectation values of the POVM elements for measurements in both the 
$H/V$ and $D/A$ bases are experimental observations. Therefore, in the 
construction of an EVM we can utilize these POVM elements. Since 
$M_H+M_V+M_{HV}+M_{\emptyset}^{+}=M_D+M_A+M_{DA}+M_{\emptyset}^{\times}=I$, 
where $I$ is the identity operator in the full state space, these POVM elements 
are not linearly independent. To construct an EVM, we will use the 
linearly independent operators in the following set:
\begin{equation}
\mathbf{\mathcal{S}}\equiv\{I, M_H, M_V, M_{HV}, M_D, M_A, M_{DA}\}. 
\label{eq:op_set}
\end{equation}
Using the EVM constructed with the operators in the above set $\mathbf{\mathcal{S}}$, 
we can verify entanglement when the observed error probability and observed 
double-click probability are low enough. To illustrate this result, in 
Sect.~\ref{toy_model} we will study a particular channel connecting Alice and Bob. 
Then, it turns out that entanglement can be verified when the depolarizing
probability $\omega$ and the multi-photon probability $p$ in the channel are low 
enough, as we will later see in Fig.~\ref{active_evm_comparison}.

\subsubsection{An improved construction of EVMs}
\label{active_operators_2}
 
To improve the results, we consider projections of the operators in the set 
$\mathbf{\mathcal{S}}$ onto various photon-number subspaces. There are two reasons 
for exploiting these projections in the construction of an EVM: First, more operator 
relationships between these projections can be exploited. Second, as shown later, the 
expectation values of these projections can be bounded from experimental observations.  
Both of these help to constrain the constructed EVM. Generally speaking, 
the higher the number of considered photon-number subspaces, the stronger the 
entanglement-verification power of our method becomes. However, with the increase 
of the number of considered photon-number subspaces the complexity of the resulting 
SDP problem increases. 

For implementation simplicity, we consider the projections of operators onto the 
zero-photon, one-photon, and two-photons subspaces. The projections 
of the identity operator $I$ onto those subspaces are denoted by $I_{1\times 1}$, 
$I_{2\times 2}$ and $I_{3\times 3}$, respectively, where $I_{d\times d}$ is the identity 
matrix of dimension $d\times d$. For the other operators in the set $\mathbf{\mathcal{S}}$, 
from their explicit expressions (e.g., Eqs.~\eqref{eq:active_HV_POVMs} and~\eqref{eq:active_DA_POVMs}) 
we can see that their projections onto a photon-number subspace are linear combinations of 
ideal operators in the same subspace, where the combination coefficients depend on mismatched 
efficiencies. Here, the ideal operators are the projections of measurement POVM elements 
as setting all detection efficiencies to be perfect; these ideal operators will be 
denoted by notations with tildes in order to be distinguished from real operators. For 
example, the projection of $M_H$ in Eq.~\eqref{eq:active_HV_POVMs} onto the ($\leq 2$)-photons 
subspace $M_H^{(\leq 2)}$ is 
\begin{align}
M_H^{(\leq 2)}= &\eta_{H/D}(1-\eta_{V/A})(I_{3\times 3}-\tilde{M}_H^{(2)}-\tilde{M}_V^{(2)}) \notag \\
&+(1-(1-\eta_{H/D})^2)\tilde{M}_H^{(2)}+\eta_{H/D} \tilde{M}_H^{(1)},
\label{eq:exp_upper_bound}
\end{align}
where the superscript `$(1)$' or `$(2)$' means restriction to the one-photon or two-photons subspace. 
To write down all the projections, we need the following set of ideal operators:
\begin{equation}
\{I_{1\times 1}, I_{2\times 2}, I_{3\times 3}, \tilde{M}_H^{(1)},\tilde{M}_D^{(1)},\tilde{M}_H^{(2)},\tilde{M}_V^{(2)},\tilde{M}_D^{(2)},\tilde{M}_A^{(2)}\}.
\label{eq:active_12photon}
\end{equation}
Note that we do not need the ideal operators $\tilde{M}_V^{(1)}, \tilde{M}_A^{(1)}, 
\tilde{M}_{HV}^{(2)}$ and $\tilde{M}_{DA}^{(2)}$, due to the linear dependences
$ \tilde{M}_H^{(1)}+\tilde{M}_V^{(1)}=\tilde{M}_D^{(1)}+\tilde M_A^{(1)}=I_{2\times 2}$ 
and $\tilde M_H^{(2)}+\tilde M_{HV}^{(2)}+\tilde M_V^{(2)}=
\tilde M_D^{(2)}+\tilde M_{DA}^{(2)}+\tilde M_A^{(2)} =I_{3\times 3}$.  
Instead of the projections onto the ($\leq2$)-photons subspace, we will use the 
ideal operators in Eq.~\eqref{eq:active_12photon} to construct an EVM, since the 
relationships between these ideal operators, as studied in Appendix~\ref{sect:spin_operators}, 
are simpler.

Moreover, there are relations between the ideal operators in Eq.~\eqref{eq:active_12photon} 
and measurement POVM elements. For example, because the POVM element $M_H$ is block-diagonal 
with respect to various photon-number subspaces, we have that 
\begin{equation}
M_H\geq M_H^{(\leq2)}.
\label{eq:exp_upper_bound2}
\end{equation}
Then, considering Eq.~\eqref{eq:exp_upper_bound}, we can relate a linear combination of ideal 
operators in Eq.~\eqref{eq:active_12photon} to the measurement POVM element $M_H$ by an 
inequality. Similar relations apply to other POVM elements. As a result, we can bound the 
expectation values of the ideal operators in Eq.~\eqref{eq:active_12photon} based on 
experimental observations. Therefore, to construct an EVM, in addition to the measurement
operators in the set defined in Eq.~\eqref{eq:op_set} we can take advantage of the ideal 
operators in the following sets: 
\begin{align}
\mathbf{\mathcal{S}}_{0}\equiv& \{\Ket{\rm{Vac}}\Bra{\rm{Vac}}\}, \label{eq:op_set0} \\
\mathbf{\mathcal{S}}_{1}\equiv& \{I_{2\times 2}, \tilde M_H^{(1)}, \tilde M_D^{(1)}, \sigma_y\},\text{ and }  \label{eq:op_set1} \\
\mathbf{\mathcal{S}}_{2}\equiv& \{I_{3\times 3}, \tilde M_H^{(2)}, \tilde M_V^{(2)}, \tilde M_D^{(2)}, \tilde M_A^{(2)}, S_y\}.
\label{eq:op_set2}
\end{align}
Here, $\Ket{\rm{Vac}}\Bra{\rm{Vac}}$ is the projection onto the vacuum 
state $\Ket{\rm{Vac}}$ (i.e., the zero-photon subspace), and the set $\mathbf{\mathcal{S}}_{1}$ 
or $\mathbf{\mathcal{S}}_{2}$ contains ideal operators in the one-photon or two-photons 
subspace, respectively. We include the qubit Pauli operator $\sigma_y$ and 
the spin-1 operator $S_y$ in the above sets, because they are involved in the 
commutation relationships between ideal operators (see Appendix~\ref{sect:spin_operators}).
Note that any two operators from any two different sets as above are orthogonal 
to each other.

\subsection{Bounds on the number of photons arriving at Bob}
\label{photon_number_bounds}
In the previous subsection, we discussed several sets of ideal operators 
exploited for constructing an EVM. From experimental observations, we can bound 
expectation values of these ideal operators. Strictly speaking, we can bound 
their expectation values only from above (see Eq.~\eqref{eq:exp_upper_bound2} for an 
example). These upper-bound constraints can be satisfied in a trivial way, if the state 
does not lie in the same Hilbert space as the ideal operators exploited and so all expectation 
values of these ideal operators are zero. As a consequence, the relationships between 
these operators would not be helpful.  Since we would like to exploit the operators 
particularly in the zero-photon, one-photon, or two-photons subspaces, we need to bound from 
below the probabilities that the state lies in these subspaces. In order to 
achieve this goal, we introduce additional constraints outside of the EVM 
formalism.

\subsubsection{Active-detection case}
For the active-detection scheme we consider the following intuition: 
With the increase of the number of photons $n$ arriving at Bob, the double-click probability (or the 
effective-error probability as defined below) conditional on the photon number $n$ will increase and
finally surpass the observed double-click probability (or the observed effective-error probability). 
Hence, in order to be consistent with experimental observations, the probability of a large number of 
photons arriving at Bob must be small. This motivates us to exploit the following double-click operator 
\begin{equation}
F_{DC}=\frac{1}{2} I^{A}\otimes M_{HV}^B+\frac{1}{2} I^{A}\otimes M_{DA}^B, 
\label{eq:active_dc_op}
\end{equation}
and the effective-error operator
\begin{align}
F_{EE}&= \frac{1}{2} M_{H}^{A}\otimes (M_{V}^{B}+\frac{1}{2}M_{HV}^{B})+\frac{1}{2} M_{V}^{A}\otimes (M_{H}^{B}+\frac{1}{2}M_{HV}^{B})\notag \\
& +\frac{1}{2} M_{D}^{A}\otimes (M_{A}^{B}+\frac{1}{2}M_{DA}^{B})+\frac{1}{2} M_{A}^{A}\otimes (M_{D}^{B}+\frac{1}{2}M_{DA}^{B}),
\label{eq:active_error_op}
\end{align}
where the superscripts `$A$' and `$B$' denote Alice and Bob.  
The coefficient $1/2$ before each term is due to the probability $1/2$ of selecting 
the $H/V$ or $D/A$ measurement basis. According to the source-replacement 
description~\cite{BBM1992}, Alice's measurement operators $M_H^A$, $M_V^A$, $M_D^A$ 
and $M_A^A$ are ideal measurement operators in the one-photon subspace. 
Bob's measurement operators are as discussed in Sect.~\ref{evm_operators}
and Appendix~\ref{active_operators_2spatial}. Note that the definition of the 
above effective-error operator is motivated by the post-processing rule 
typically used in squashing models~\cite{normand2008, Kiyoshi2008}, where 
one uniformly randomly assigns a double-click event to one of the two 
single-click events at the same basis. 

Before explaining how to utilize the operators $F_{DC}$ and $F_{EE}$, let us
discuss two properties of the state $\rho_{AB}$ shared between Alice 
and Bob in the thought setup according to the source-replacement description. 
First, because measurement POVMs at Bob are block-diagonal 
with respect to various photon-number subspaces across all modes involved,  
we can assume without loss of generality that the state $\rho_{AB}$ 
has the same block-diagonal structure. That is, $\rho_{AB}$ 
can be written as 
\begin{equation}
\rho_{AB}=\bigoplus_{n=0}^{\infty}p_n\rho_{AB}^{(n)}.
\label{rho_block}
\end{equation}
Here, $p_n$ is the probability that the state $\rho_{AB}$ lies in the $n$-photons 
subspace across all modes involved, and $\rho_{AB}^{(n)}$ is the normalized state 
conditional on $n$ photons arriving  at Bob. Second, we can assume without loss 
of generality that $\rho_{AB}$ and $\rho_{AB}^{(n)}$ are real-valued. This is 
because all measurement POVM elements $M_x^A$ and $M_y^B$ of Alice and Bob can be 
represented by real-valued matrices in the photon-number basis 
(see Ref.~\cite{Johannes2006} for a detailed argument). 
The above two properties apply to the state $\rho_{AB}$ under either the active- 
or passive-detection scheme. Considering the second property, we can reduce the number 
of free parameters in the constructed EVM and in the following optimization problems. 
 
Now, let us proceed to formalize our intuition that the double-click probability 
(or the effective-error probability) conditional on the number of photons
arriving at Bob increases with the photon number $n$.  
We study the following optimization problems:
\begin{equation}
\label{eq:min_dc}
\begin{array}{rc}
d_{n,min}=& \min_{\rho_{AB}^{(n)}} d_n \\
 \text{subject to}&  \rho_{AB}^{(n)}\geq 0   \\
& \Tr\left(\rho_{AB}^{(n)}\right)=1  \\
& \left(\rho_{AB}^{(n)}\right)^{\Gamma_A} \geq 0 
\end{array}
\end{equation}
and 
\begin{equation}
\label{eq:min_error}
\begin{array}{rc}
e_{n,min}=& \min_{\rho_{AB}^{(n)}} e_n \\
 \text{subject to}&  \rho_{AB}^{(n)}\geq 0   \\
& \Tr\left(\rho_{AB}^{(n)}\right)=1  \\
& \left(\rho_{AB}^{(n)}\right)^{\Gamma_A} \geq 0, 
\end{array}
\end{equation}
where 
the objective functions of the above two optimization problems 
are given by $d_n=\Tr\left(\rho_{AB}^{(n)}F_{DC}^{(n)}\right)$ and
$e_n=\Tr\left(\rho_{AB}^{(n)}F_{EE}^{(n)}\right)$, respectively. The operators
$F_{DC}^{(n)}$ and $F_{EE}^{(n)}$ are projections of the double-click operator $F_{DC}$
and the effective-error operator $F_{EE}$ onto the $n$-photons subspace of Bob. 
Note that in the above optimization problems we constrain the $n$-photons state 
$\rho_{AB}^{(n)}$ such that its partial transpose $\left(\rho_{AB}^{(n)}\right)^{\Gamma_A}$ 
is positive-semidefinite.  The reason is as follows: In order to verify entanglement  
we need to check whether or not there is a separable state consistent with experimental 
observations. Considering the block-diagonal structure of the state $\rho_{AB}$,
if $\rho_{AB}$ is separable then its projection onto any $n$-photons subspace
$\rho_{AB}^{(n)}$ must be separable and so satisfy the PPT criterion.  

\begin{figure}[htb!]
\includegraphics[scale=0.45,viewport=3cm 8cm 19cm 20cm]{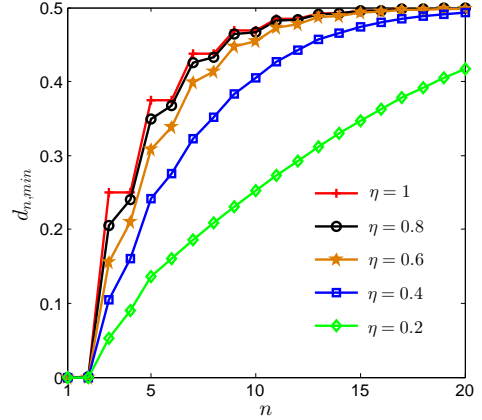}  
\caption{The minimum double-click probability $d_{n,min}$ as a function of the number of
photons $n$ arriving at Bob in the active-detection scheme. The results for 
the mismatch model specified in Table~\ref{active_mismatch} are shown.}
   \label{active_doubleclick} 
\end{figure}

\begin{figure}[htb!]
\includegraphics[scale=0.45,viewport=3cm 8.5cm 19cm 19cm]{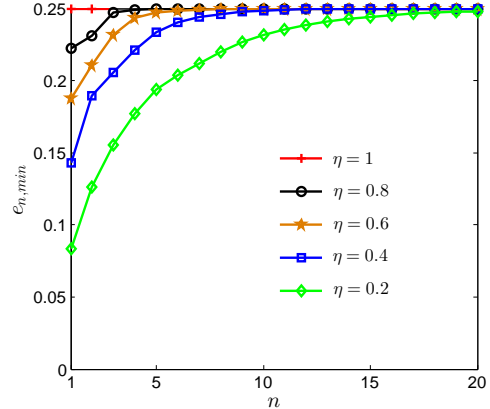}  
\caption{The minimum effective-error probability $e_{n,min}$ as a function of the 
number of photons $n$ arriving at Bob in the active-detection scheme. The results for 
the mismatch model specified in Table~\ref{active_mismatch} are shown.}
   \label{active_error} 
\end{figure}

The optimization problems in Eqs.~\eqref{eq:min_dc} and~\eqref{eq:min_error} are SDPs.
We solved them numerically using the toolbox YALMIP~\cite{YALMIP} of MATLAB. We observed 
that with the increase of the photon number $n$, both of the minimum 
double-click probability $d_{n,min}$ and the minimum effective-error probability $e_{n,min}$ 
monotonically increase and converge under an arbitrary mismatch model. 
(We have not proved this analytically, but 
numerical evidence strongly suggests that our observation is true.) The optimization results
for the mismatch model specified in Table~\ref{active_mismatch} are shown in Figs.~\ref{active_doubleclick} 
and~\ref{active_error}. As a consequence, given the observed double-click probability 
$d_\text{obs}$ and observed effective-error probability $e_\text{obs}$, we get that
\begin{align}
&d_\text{obs}=\sum_{n=0}^{\infty}p_nd_n \geq p_2d_2+(1-p_0-p_1-p_2)d_{3,min}, \label{eq:actual_doubleclick_bound} \\
& \text{and} \notag \\ 
&e_\text{obs}=\sum_{n=0}^{\infty}p_ne_n \geq p_1e_1+p_2e_2+(1-p_0-p_1-p_2)e_{3,min},
\label{eq:actual_error_bound}
\end{align}
if the state $\rho_{AB}$ is separable. Here, we use the facts that $\sum_{n=0}^{\infty}p_n=1$
and $d_0=d_1=e_0=0$. From the above two inequalities, obviously we get that $d_\text{obs}\geq (1-p_0-p_1-p_2)d_{3,min}$
and $e_\text{obs}\geq (1-p_0-p_1-p_2)e_{3,min}$. Hence, we can bound from below the sum of the 
probabilities of zero photon, one photon and two photons $(p_0+p_1+p_2)$. 
Note that the parameters $p_0$, $p_1$, $p_2$, $d_2$, $e_1$, and $e_2$ can
be written as linear combinations of the EVM entries, when the EVM is constructed 
with ideal operators in the $(\leq 2)$-photons subspace (as we implemented). 

\subsubsection{Passive-detection case}
In the passive-detection scheme as shown in Fig.~\ref{setup}(b), 
a 50/50 beam splitter is used for selecting different 
measurement bases. So, the probability that $n$ incoming photons leave the 
beam splitter at different output arms is $1-2^{-(n-1)}$, which increases 
with the photon number $n$. These photons will potentially contribute to 
clicks at two or more detectors at different output arms of the beam 
splitter, to which we refer as cross-click events. So, we expect 
that with the increase of $n$ the cross-click probability conditional on 
$n$ increases and converges to one. (In contrast, neither the double-click 
probability nor the effective-error probability increases with $n$, since 
the probability that $n$ incoming photons leave the beam splitter 
at the same output arm decreases with $n$.) The above expectation 
motivates us to consider the following cross-click operator:
\begin{equation}
F_{CC}=I^A\otimes M_{CC}^B,
\end{equation}
where $M_{CC}$ is the POVM element for cross-click events at Bob (see 
Appendix~\ref{sect:povm_with_mismatch} for details of the measurement POVM elements 
in the passive-detection scheme). Our expectation can be formalized as  
investigating the following optimization problem:
\begin{equation}
\label{eq:min_cc}
\begin{array}{rc}
c_{n,min}=& \min_{\rho_{AB}^{(n)}} c_n \\
 \text{subject to}&  \rho_{AB}^{(n)}\geq 0   \\
& \Tr\left(\rho_{AB}^{(n)}\right)=1  \\
& \left(\rho_{AB}^{(n)}\right)^{\Gamma_A} \geq 0, 
\end{array}
\end{equation}
where the objective function is given by $c_n= \Tr\left(\rho_{AB}^{(n)}F_{CC}^{(n)}\right)$, 
and $F_{CC}^{(n)}$ is the projection of the cross-click operator $F_{CC}$
onto the $n$-photons subspace of Bob. The same as in the active-detection 
case, we constrain the $n$-photons state $\rho_{AB}^{(n)}$ such that it 
satisfies the PPT criterion. 

\begin{figure}[htb!]
\includegraphics[scale=0.45,viewport=3cm 8cm 19cm 20cm]{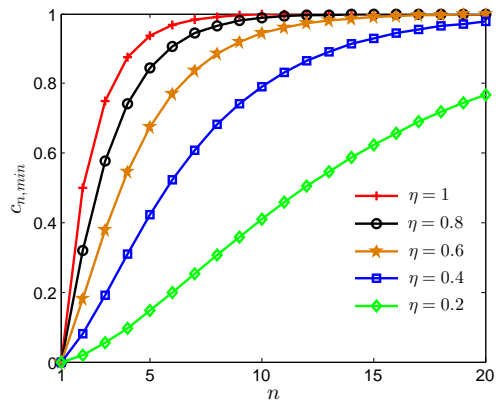}  
\caption{The minimum cross-click probability $c_{n,min}$ as a function of the number of
photons $n$ arriving at Bob in the passive-detection scheme. The results for 
the mismatch model specified in Table~\ref{passive_mismatch} are shown.}
   \label{passive_crossclick} 
\end{figure}

The above optimization problem is a SDP, which can be solved numerically 
using the toolbox YALMIP~\cite{YALMIP} of MATLAB. Strong numerical evidence 
suggests that with the increase of $n$ the minimum cross-click probability $c_{n,min}$
monotonically increases and converges to one under an arbitrary mismatch model. 
For the mismatch model specified in Table~\ref{passive_mismatch}, the 
optimization results are shown in Fig.~\ref{passive_crossclick}. As a
result, given the observed cross-click probability $c_{\rm{obs}}$, we have that
\begin{equation}
c_{\text{obs}}=\sum_{n=0}^{\infty} p_n c_n= \sum_{n=2}^{\infty} p_n c_n\geq \left( 1-p_0-p_1 \right)c_{2,min}.
\label{eq:actual_cross_bound}
\end{equation}
Here, we use the facts that $\sum_{n=0}^{\infty}p_n=1$ and $c_0=c_1=0$. 
Thus we can bound from below the probability of no more than one photon arriving at Bob. 

In the end, we would like to make a comment on the effect of the detection-efficiency  
mismatch. From Figs.~\ref{active_doubleclick},~\ref{active_error} and~\ref{passive_crossclick},
one can see that, given the photon number $n$, the larger the efficiency 
mismatch (i.e., the smaller the parameter $\eta$), the smaller the 
minimum probability of induced double-click events, effective-error events,
or cross-click events becomes. In this sense, we expect that the 
efficiency mismatch helps Eve to attack the QKD system.

\subsection{Efficiency renormalization}
\label{renormalization}
\begin{figure}[htb!]
\includegraphics [scale=0.65,viewport=8cm 15cm 16cm 19cm]{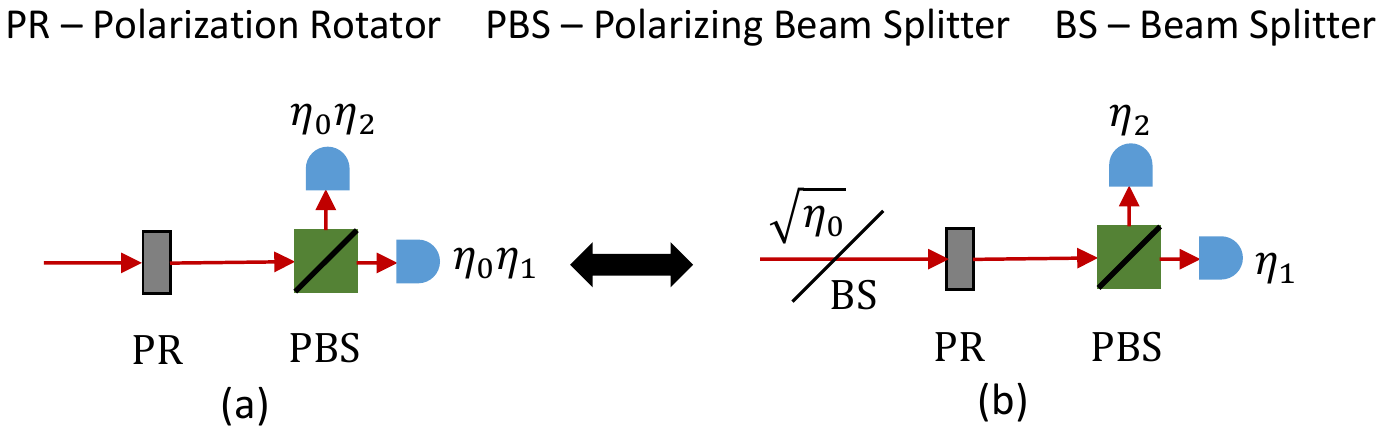}  
\caption{Two equivalent descriptions of the same measurement device in the 
active-detection scheme, where the detection efficiency is written 
down around each detector and $0\leq \eta_0, \eta_1, \eta_2\leq 1$: (a) is 
the actual situation, and (b) is the 
hypothetical situation where the common loss $\eta_0$ in the two detectors 
is factored out and treated as a part of transmission loss. The beam splitter 
in (b) has a transmission coefficient $\sqrt{\eta_0}$. See 
Appendix~\ref{sect:proof_active_equiva} for a proof of this equivalence.}
\label{active_equivalent} 
\end{figure}

The relative efficiencies of different detectors in Fig.~\ref{setup}
can be those described in Tables~\ref{active_mismatch} and~\ref{passive_mismatch}.
However, in practice no detector has an absolute efficiency $1$. Suppose that 
the efficiencies of the two detectors in the active-detection scheme of 
Fig.~\ref{setup}(a) can be written as $\eta_0\eta_1$ and $\eta_0\eta_2$ 
respectively, where $0\leq \eta_0, \eta_1, \eta_2\leq 1$. Then, there are 
two equivalent descriptions of the same 
measurement device, as shown in Fig.~\ref{active_equivalent}. According 
to the description in Fig.~\ref{active_equivalent}(b), we can lump together 
the common loss $\eta_0$ in the two detectors and the transmission loss 
in order to verify entanglement. The reason is as follows: Once 
entanglement is verified in the state shared between Alice and Bob after 
passing the beam splitter in Fig.~\ref{active_equivalent}(b), then the state 
before entering the whole measurement device in Fig.~\ref{active_equivalent}(a) 
must be entangled (otherwise, there is contradiction).
 
\begin{figure}[htb!]
\includegraphics [scale=0.63,viewport=5cm 17.8cm 16cm 24cm]{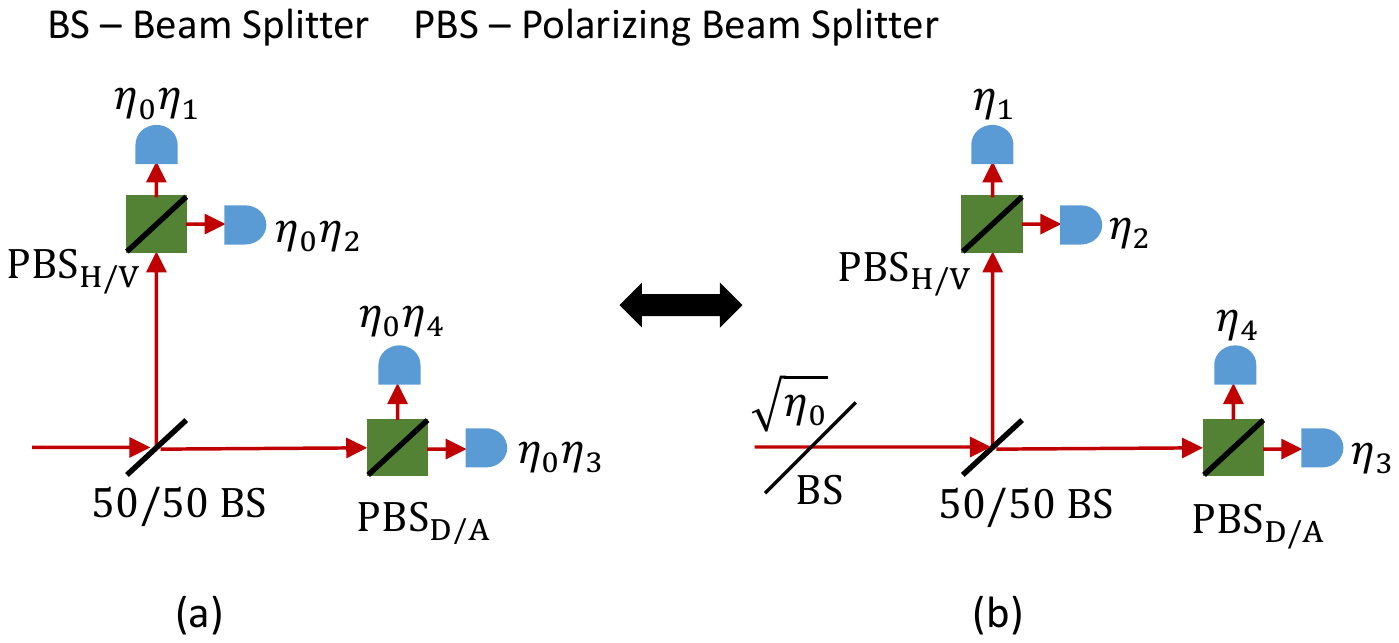}  
\caption{Two equivalent descriptions of the same measurement device in the 
passive-detection scheme, where the detection efficiency is written 
down around each detector and $0\leq \eta_0, \eta_1, \eta_2, \eta_3, \eta_4 \leq 1$: 
(a) is the actual situation, and (b) is the 
hypothetical situation where the common loss $\eta_0$ in the four detectors 
is factored out and treated as a part of transmission loss. The first beam 
splitter in (b) has a transmission coefficient $\sqrt{\eta_0}$.  
See Appendix~\ref{sect:proof_passive_equiva} for a proof of this equivalence. 
}
   \label{passive_equivalent} 
\end{figure}

According to Fig.~\ref{active_equivalent}(b), the detection efficiencies 
$\eta_{H/D}$ and $\eta_{V/A}$ in measurement POVM elements, such as those 
in Eqs.~\eqref{eq:active_HV_POVMs} and \eqref{eq:active_DA_POVMs}, are  
rescaled by a factor $1/\eta_0$. As we observed, given the photon 
number $n$ arriving at Bob, with the increase of these detection efficiencies 
both the minimum double-click probability $d_{n,min}$ and the minimum 
effective-error probability $e_{n,min}$ increase. Hence, given experimental 
observations, by rescaling detection efficiencies the expectation values of 
ideal operators in the ($\leq 2$)-photons subspace can be bounded more tightly. 
We observed that more entangled states can be verified in this way than according 
to the description in Fig.~\ref{active_equivalent}(a). Therefore, in 
implementations of our method we renormalize detection efficiencies so that 
the maximum relative efficiency is $1$. The same trick can be applied to the 
passive-detection scheme, as shown in Fig.~\ref{passive_equivalent}.  
Note that the renormalization trick can also be applied to the case with 
multiple spatial modes. But, for this general case we need to renormalize 
the efficiencies over all detectors and over all spatial modes at the same 
time.  

In a nutshell, our numerical observations show that if we renormalize detection 
efficiencies so that the maximum relative efficiency over all detectors and over 
all spatial modes is $1$, then we can verify more entangled states than according 
to the actual description of the measurement device.   

\subsection{SDP for entanglement verification}
\label{sdp}
According to \textbf{Observation~\ref{evm_ppt}}, we can formulate entanglement verification as 
a SDP problem. Specifically, we need to solve a SDP feasibility problem
of the form
\begin{equation}
\label{eq:sdp}
\begin{array}{rc}
\text{find}& \chi \\
 \text{subject to}&  \chi\geq 0 \text{  and  }  \chi^{\Gamma_A}\geq 0  \\
& \sum_{ijkl}C^{(n)}_{ij,kl} \chi_{ij,kl}= 0, n=1,2,...,N \\
& \sum_{ijkl}C^{(m)}_{ij,kl} \chi_{ij,kl} \geq 0, m=N+1,2,...,N+M.
\end{array}
\end{equation}
Here, the matrix $\chi$ is an EVM, the coefficients $C^{(n)}_{ij,kl}$ 
with $n=1,2,...,N+M$ are complex numbers, and $N$ and $M$ are the numbers 
of equality and inequality constraints respectively.  
The equality constraints can be according to experimental observations and 
commutation relationships between operators (such as those in 
Eq.~\eqref{eq:active_commutation} of Appendix~\ref{sect:spin_operators}). 
The inequality constraints can be derived from operator relationships, such as 
those in Eqs.~\eqref{eq:active_HV_rela} and~\eqref{eq:exp_upper_bound2}, 
or based on the inequalities in Eqs.~\eqref{eq:actual_doubleclick_bound},
\eqref{eq:actual_error_bound}, and ~\eqref{eq:actual_cross_bound}. If the 
SDP problem in Eq.~\eqref{eq:sdp} is not feasible, then the underlying state 
shared by Alice and Bob must be entangled. In our implementation, the 
optimization problem in Eq.~\eqref{eq:sdp} is solved using the toolbox 
YALMIP~\cite{YALMIP} of MATLAB.  More details on the formulation
of the SDP problem can be found in Appendix~\ref{sect:sdp_details}.

\section{Demonstration with simulated results} 
\label{results}
The method discussed in Sect.~\ref{method} works for general detection-efficiency
mismatch models. To illustrate our method, we consider particular mismatch 
models, such as those specified in Tables~\ref{active_mismatch} and \ref{passive_mismatch}.
In the absence of a real experiment and for simplicity, we simulate experimental 
results according to a toy model detailed in the following subsection. We would like to 
stress that our method for verifying entanglement depends only on experimental observations 
and measurement POVMs but does not depend on the details of data simulation listed below.

\subsection{Data simulation}
\label{toy_model}
We consider the ideal BB84-QKD protocol. Alice first prepares an entangled state 
as in Eq.~\eqref{eq:ent_BB84} according to the source-replacement description.
Then Alice sends out the polarized photon to Bob. We model the channel connecting 
Alice and Bob as a depolarizing channel with depolarizing probability $\omega$; 
additionally, the transmission loss (i.e., the single-photon loss probability) 
over the channel is $r$; and Eve intercepts the single photon and resends multiple 
photons to Bob with probability $p$. The multi-photon state resent by Eve is a 
randomly-polarized $n$-photons state of the form
\begin{equation}
\rho_n=\frac{1}{2n\pi}\int_0^{2\pi} \mathrm{d}\theta \left(\hat{a}_\theta^\dagger\right)^n \ket{0}\bra{0} \left(\hat{a}_\theta\right)^n.
\label{sim_state}
\end{equation}
Here, $\hat{a}_\theta^\dagger=\cos(\theta) \hat{a}_H^\dagger+\sin(\theta)\hat{a}_V^\dagger$, 
and $\hat{a}_H^\dagger$ (or $\hat{a}_V^\dagger$) is the creation operator
associated with the horizontally-polarized (or vertically-polarized) optical 
mode. The number of photons $n$ resent will be specified later for each 
simulation. Moreover, when there is more than one spatial mode, the 
optical signal is distributed over these spatial modes uniformly randomly. 
For measurements at Bob, we consider both the active- 
and passive-detection schemes, as shown in Fig.~\ref{setup}.

\subsection{No-mismatch case and comparison with squashing models}
When there is no efficiency mismatch between threshold detectors involved 
in the measurement device, one can verify entanglement based on squashing 
models~\cite{normand2008, Kiyoshi2008, Moroder2010, Gittsovich2014}.  
In this case, Bob uniformly randomly assigns a double-click event to 
one of the two single-click events at the same measurement basis, and 
assigns all cross-click events to no detection. Then, his observations 
can be thought of as generated by a qutrit system (constituted by a single 
photon and the vacuum). Once we can verify that this qutrit system is 
entangled with Alice's system, then the original system shared by Alice and 
Bob must be entangled~\cite{normand2008,Moroder2010}. Note that it is 
still an open question whether or not a squashing model exists in the case 
of efficiency mismatch.

\begin{figure}[htb!]
\includegraphics[scale=0.5,viewport=3cm 8.3cm 19cm 19.8cm]{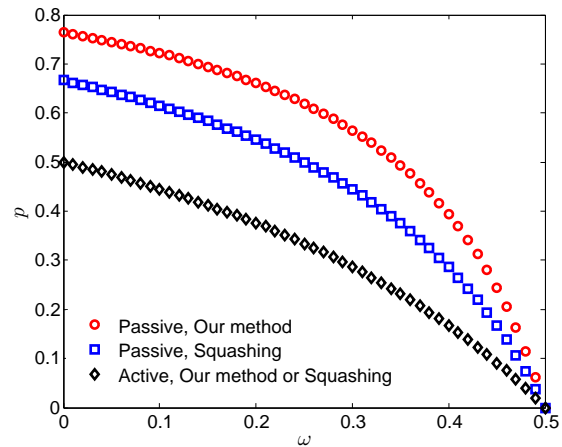}  
\caption{Comparison of our method with the one based on squashing models 
when there is no efficiency mismatch and no transmission loss. Here,
$\omega$ is the depolarizing probability and $p$ is the multi-photon 
probability in the channel. Below the curves entanglement can be verified. 
For the active-detection scheme, the results according to our method or 
based on the squashing model coincide with each other, as shown by diamonds. 
For the passive-detection scheme, the results according to our method are 
shown by circles, whereas the results based on the squashing model are shown 
by squares.}
\label{squash} 
\end{figure}

We compare our method with the one based on squashing models when there is 
no efficiency mismatch. For this purpose, we simulate experimental results
according to the toy model specified in Sec.~\ref{toy_model}. Particularly,
we consider the case with no transmission loss. The comparison shows different 
behaviour, depending on the observations of Alice and Bob. To demonstrate 
the advantage of our method, we consider the case that the multi-photon
state resent by Eve is of the form as in Eq.~\eqref{sim_state} with $n=2$. 
(Note that the advantage of our method reduces with the increase of $n$ and 
finally disappears.) To construct an EVM, we use both the measurement 
POVM elements (Eq.~\eqref{eq:op_set} for the active-detection scheme or
Eq.~\eqref{eq:passive_op_set} in Appendix~\ref{passive_operators} for 
the passive-detection scheme) and the ideal operators in the ($\leq 2$)-photons 
subspace (Eqs.~\eqref{eq:op_set0},~\eqref{eq:op_set1} and~\eqref{eq:op_set2}).
The results in Fig.~\ref{squash} demonstrate the advantage of our method for 
entanglement verification in the passive-detection scheme. This could be 
understood as follows: According to the squashing model we discard cross-click 
events, whereas in our method we take advantage of them in order to bound the 
number of photons arriving at Bob (see Eq.~\eqref{eq:actual_cross_bound}). 
Fig.~\ref{squash} also shows that when there is no mismatch the passive-detection 
scheme is better for verifying entanglement than the active-detection scheme. 
This is because the probability of detecting multi-photon events in the 
passive-detection scheme is higher than that in the active-detection scheme, 
given the same incoming multi-photon state. So, in the passive-detection 
scheme our method can take more advantage of operators in the $(\leq 2)$-photons 
subspace.

\begin{figure}[htb!]
\includegraphics[scale=0.5,viewport=3cm 9cm 19cm 19cm]{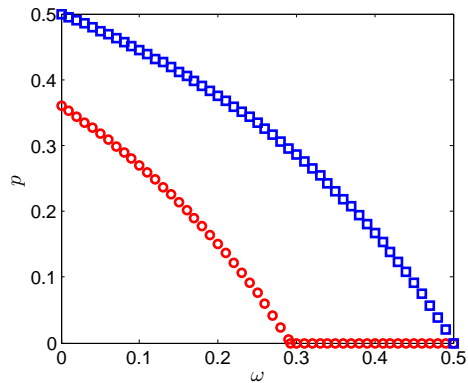}  
\caption{Comparison of the results using EVMs constructed with different
numbers of operators in the active-detection scheme.  Here, $\omega$ is the 
depolarizing probability and $p$ is the multi-photon probability in the 
channel. Below the curves entanglement can be verified. The circles show the 
results when using only the measurement POVM elements (Eq.~\eqref{eq:op_set}) 
to construct an EVM, whereas the squares show the results when using both the 
measurement POVM elements (Eq.~\eqref{eq:op_set}) and the ideal operators in the 
($\leq 2$)-photons subspace (Eqs.~\eqref{eq:op_set0},~\eqref{eq:op_set1}
and~\eqref{eq:op_set2}) to construct an EVM. Here, we consider the case with no 
efficiency mismatch and no transmission loss.}
\label{active_evm_comparison} 
\end{figure}

We also observe that it is useful to consider operators in various photon-number 
subspaces, as demonstrated in Fig.~\ref{active_evm_comparison}. In general, the 
higher the number of considered photon-number subspaces, the stronger the 
entanglement-verification power of our method becomes. But, with the increase 
of the number of considered photon-number subspaces the complexity 
of the resulting SDP problem increases.

\subsection{Spatial-mode-independent mismatch} 
Let us start by considering the simple case where the efficiencies of 
various detectors involved can take different values, but the efficiencies 
are equal for every spatial mode. Without loss of generality, we assume 
that there is only one spatial mode. To construct an EVM, we use both the 
measurement POVM elements (Eq.~\eqref{eq:op_set} for the active-detection 
scheme or Eq.~\eqref{eq:passive_op_set} in Appendix~\ref{passive_operators}
for the passive-detection scheme) and the ideal operators in the ($\leq 2$)-photons 
subspace (Eqs.~\eqref{eq:op_set0},~\eqref{eq:op_set1} and~\eqref{eq:op_set2}). 
The results presented in this subsection are based on simulations according 
to the toy model specified in Sec.~\ref{toy_model}. For simplicity, we assume   
that the multi-photon state resent by Eve is of the form as in 
Eq.~\eqref{sim_state} with $n\rightarrow\infty$.

\begin{figure}[htb!]
\includegraphics[scale=0.5,viewport=4cm 8.6cm 18.5cm 19cm]{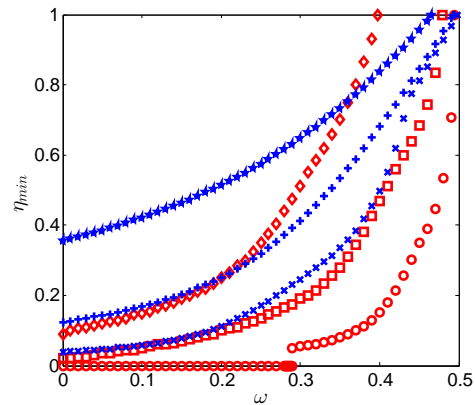}  
\caption{Minimum efficiency $\eta_{min}$ required for verifying entanglement as
a function of the depolarizing probability $\omega$ in the channel. The 
circles, squares and diamonds show the results for the active-detection scheme
with transmission losses $r=0, 0.75$ and $0.95$, respectively. The crosses,
pluses and pentagrams show the results for the passive-detection scheme with 
transmission losses $r=0, 0.75$ and $0.95$, respectively. Here, we fix the  
multi-photon probability $p=0.01$. We choose the above values of $r$  
and $p$ just for ease of graphical illustrations.}
\label{com_1spatial} 
\end{figure}

First, we compare the abilities of verifying entanglement of the two detection 
schemes in Fig.~\ref{setup}, when efficiency mismatch is the same. Since there 
are two detectors in the active-detection scheme, up to permutation symmetry 
and efficiency renormalization as discussed in Sect.~\ref{renormalization}, 
there is only one kind of mismatch model. Hence, for the purpose of comparison, 
we consider the case $\eta_{H/D}=1$ and $\eta_{V/A}=\eta<1$ in the 
active-detection scheme, corresponding to the case $\eta_H=\eta_D=1$ and 
$\eta_V=\eta_A=\eta<1$ in the passive-detection scheme. We would like to find 
out the minimum efficiency $\eta_{min}$ (corresponding to the maximum mismatch) 
such that entanglement can be verified as long as $\eta\geq \eta_{min}$. This 
minimum efficiency $\eta_{min}$ characterizes the robustness of a detection 
scheme against mismatch for verifying entanglement.  Typical results are 
shown in Fig.~\ref{com_1spatial}, where we fix the multi-photon probability $p$ 
and the transmission loss $r$, and characterize the minimum efficiency $\eta_{min}$  
as a function of the depolarizing probability $\omega$. From Fig.~\ref{com_1spatial},   
one can see that the minimum efficiencies in the active- and passive-detection schemes  
cross over with each other: For most values of $\omega$ the active-detection scheme  
is better than the passive-detection scheme in terms of robustness against mismatch.  
However, there are regions for the values of $\omega$ where the passive-detection  
scheme is better. Fig.~\ref{com_1spatial} also shows that our method works well
even for high-loss cases.

When there are no multi-photon events, i.e., $p=0$, our method can verify 
entanglement as long as the depolarizing probability satisfies $\omega<1/2$, 
regardless of the values for the detection efficiency $\eta$ and transmission 
loss $r$. This is because when $p=0$ our method can certify that there is no 
more than one photon arriving at Bob. Then, given the observed probability distribution 
and the mismatch model, the probability distribution for the no-mismatch case can 
be inferred. In the case of no mismatch, entanglement can be verified as long 
as the quantum bit error rate is less than $25\%$~\cite{qkdreview2002}, which  
corresponds to the condition that the depolarizing probability satisfies $\omega<1/2$.

\begin{figure}[htb!]
\includegraphics[scale=0.5,viewport=3cm 8.5cm 19cm 19.2cm]{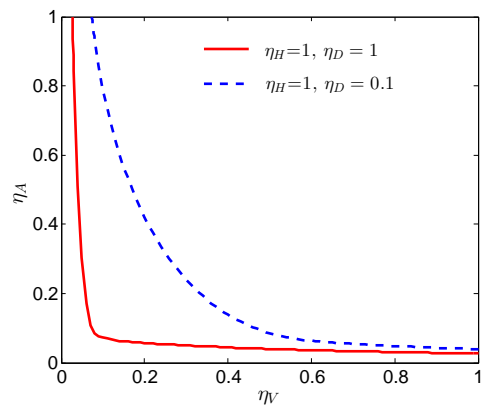}  
\caption{Trade-off between efficiencies $\eta_V$ and $\eta_A$ under fixed 
values for $\eta_H$ and $\eta_D$ in the passive-detection scheme. When the 
point $(\eta_V, \eta_A)$ is above the curve, entanglement can be verified
by our method. Here, we fix the depolarizing probability $\omega=0.05$, 
the multi-photon probability $p=0.01$, and the transmission loss $r=0.5$. 
We choose these values just for ease of graphical illustrations.}
\label{tradeoff_passive_1spatial} 
\end{figure}

Second, we study more general mismatch models in the passive-detection scheme. 
In this scheme, the efficiencies of the four detectors as shown in Fig.~\ref{setup}(b) 
can take different values from each other. When fixing the efficiencies of 
two of the four detectors, for example, $\eta_H$ and $\eta_D$, there is a 
trade-off between the efficiencies of the other two detectors, $\eta_V$ and 
$\eta_A$, in order to verify entanglement, as shown in Fig.~\ref{tradeoff_passive_1spatial}.

\subsection{Spatial-mode-dependent mismatch} 
We now increase the number of spatial modes and consider spatial-mode-dependent  
mismatch models, such as those in Tables~\ref{active_mismatch} and~\ref{passive_mismatch}. 
The results presented in this subsection are based on simulations according to the  
toy model specified in Sec.~\ref{toy_model}. As in the previous subsection, we  
assume that the multi-photon state resent by Eve is of the form as in  
Eq.~\eqref{sim_state} with $n\rightarrow\infty$.

\begin{figure}[htb!]
\includegraphics[scale=0.5,viewport=3cm 8.8cm 19cm 19.3cm]{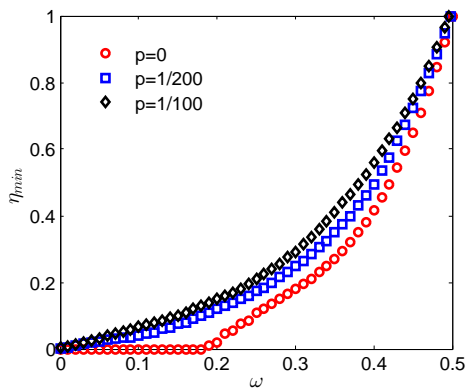} 
\caption{Minimum efficiency $\eta_{min}$ required for verifying entanglement as
a function of the depolarizing probability $\omega$ in the channel without 
transmission loss. Different markers are for different multi-photon probabilities 
$p$ as labelled in the plot. Here we consider the active-detection scheme.
The mismatch model studied is shown in Table~\ref{active_mismatch}. We  
choose the above values of $p$ just for ease of graphical illustrations.} 
\label{active_noloss}
\end{figure}

\begin{figure}
\includegraphics[scale=0.5,viewport=3cm 8.5cm 19cm 19.3cm]{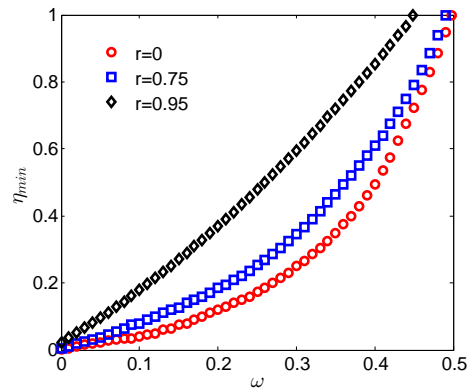}
\caption{Minimum efficiency $\eta_{min}$ required for verifying entanglement 
as a function of the depolarizing probability $\omega$ in the channel
with transmission loss. Different markers are for different transmission 
losses $r$ as labelled in the plot. Here we consider the active-detection 
scheme and fix the multi-photon probability $p=1/200$.  The mismatch model 
studied is shown in Table~\ref{active_mismatch}. We choose the above values 
of $r$ and $p$ just for ease of graphical illustrations.} 
\label{active_loss}  
\end{figure}

First, let us study the mismatch model in Table~\ref{active_mismatch} for the 
active-detection scheme. As in the previous subsection, we would like to find out 
the minimum efficiency $\eta_{min}$ characterizing the robustness of a detection 
scheme against mismatch for verifying entanglement. Here, we use both  
measurement POVM elements and the ideal operators in the ($\leq 2$)-photons  
subspace to construct an EVM (see Appendix~\ref{active_operators_2spatial} for 
details). When there is no transmission loss, i.e., $r=0$, the results are shown 
in Fig.~\ref{active_noloss}. From this figure, one can see that the higher the 
depolarizing probability $\omega$ or the multi-photon probability $p$, 
the larger the minimum efficiency $\eta_{min}$ becomes for verifying entanglement. 
We also study the effect of transmission loss on entanglement verification, as  
shown in Fig.~\ref{active_loss}. From this figure, one can see that our method
works well even for high-loss cases. The results in Figs.~\ref{active_noloss}
and~\ref{active_loss} suggest that the larger the efficiency mismatch
(i.e., the smaller the efficiency $\eta$), the smaller the set of noise parameters 
$\omega$, $p$ and $r$ that preserve entanglement becomes.  

\begin{figure}
\includegraphics[scale=0.5,viewport=3cm 8.8cm 19cm 19.8cm]{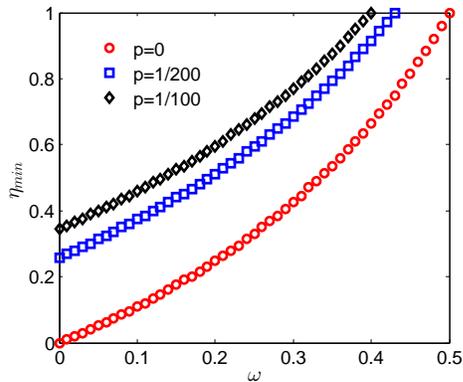} 
\caption{Minimum efficiency $\eta_{min}$ required for verifying entanglement as
a function of the depolarizing probability $\omega$ in the channel without
transmission loss. Different markers are for different multi-photon probabilities 
$p$ as labelled in the plot. Here we consider the passive-detection scheme. 
The mismatch model studied is shown in Table~\ref{passive_mismatch}. We choose 
the above values of $p$ just for ease of graphical illustrations.}
\label{passive_noloss} 
\end{figure}

\begin{figure}
\includegraphics[scale=0.5,viewport=3cm 8.5cm 19cm 19.5cm]{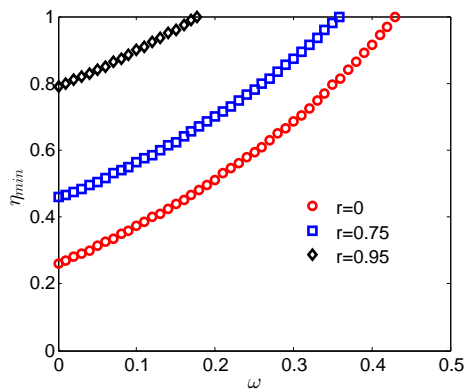} 
\caption{Minimum efficiency $\eta_{min}$ required for verifying entanglement  
as a function of the depolarizing probability $\omega$ in the channel with
transmission loss. Different markers are for different transmission losses 
$r$ as labelled in the plot. Here we consider the passive-detection scheme 
and fix the multi-photon probability $p=1/200$.  The mismatch model studied 
is shown in Table~\ref{passive_mismatch}. We choose the above values of $r$ 
and  $p$ just for ease of graphical illustrations.}
\label{passive_loss} 
\end{figure}

Second, we study the mismatch model in Table~\ref{passive_mismatch} for 
the passive-detection scheme. Here, we use both measurement POVM elements 
and the ideal operators in the ($\leq 1$)-photon subspace to construct an EVM.
Because of the implementation complexity, we do not consider the operators in 
the two-photons subspace (see Appendix~\ref{passive_operators} for details).
The results with or without transmission loss are shown in Figs.~\ref{passive_noloss} 
and \ref{passive_loss}, respectively. These results also suggest that the larger 
the efficiency mismatch (i.e., the smaller the efficiency $\eta$), the smaller 
the set of noise parameters $\omega$, $p$ and $r$ that preserve entanglement becomes.  

Note that we cannot compare the robustness of the two detection schemes against 
mismatch for verifying entanglement via Figs.~\ref{active_noloss} and~\ref{passive_noloss} 
or via Figs.\ref{active_loss} and~\ref{passive_loss}. The reasons are as follows: 
First, the mismatch models studied in Tables~\ref{active_mismatch} and~\ref{passive_mismatch}, 
for the active- and passive-detection schemes respectively, are different. 
There is no one-to-one correspondence between these two mismatch models. Second, 
the EVMs for different detection schemes are constructed using different sets of 
operators. For the active-detection scheme we use the ideal operators in the 
($\leq 2$)-photons subspace, whereas for the passive-detection scheme we use the 
ideal operators only in the ($\leq 1$)-photon subspace. The higher the number 
of considered photon-number subspaces, the stronger the entanglement-verification 
power of our method becomes. Hence, the comparison of the two detection schemes 
via Figs.~\ref{active_noloss} and~\ref{passive_noloss} or via Figs.~\ref{active_loss} 
and~\ref{passive_loss} would not be fair.  We would like to stress that we have  
developed a general method for verifying entanglement with efficiency mismatch.  
How to optimize our method and improve its entanglement-verification power will 
require future study.

In the end, we would like to make two notes. First, numerical results
suggest that when there are no multi-photon events the ability 
of our method to verify entanglement does not depend on transmission 
loss. (The results without multi-photon events and without
transmission loss are shown in Figs.~\ref{active_noloss}  
and~\ref{passive_noloss}, for the active- and passive-detection 
schemes respectively.) Second, we studied the efficiency mismatch 
in the experiment of Ref.~\cite{Shihan2015}. As Ref.~\cite{Shihan2015} 
demonstrated, not only is efficiency mismatched between the four 
detectors used in the passive-detection scheme, but also the mismatch 
depends on which one of the four spatial modes contains the incoming 
optical signal.  Ref.~\cite{Shihan2015} studied two different cases,  
i.e., with or without a pinhole inserting in front of the measurement  
device. When there is no pinhole, the observed mismatch, as shown in 
Table~\ref{passive_mismatch_nopinhole}, is so large that successful  
intercept-resend attacks on the QKD system exist~\cite{Shihan2015}. 
When there is a pinhole, the observed mismatch, as shown in  
Table~\ref{passive_mismatch_pinhole}, is reduced so that our method 
can be used to verify entanglement.

\begin{table}
\caption{Spatial-mode-dependent mismatch observed in Ref.~\cite{Shihan2015}, 
when there is no pinhole. Different columns are for different detectors labelled 
and shown in Fig.~\ref{setup}(b). Different rows are mismatched efficiencies 
for different spatial modes.}
\begin{center}
\begin{tabular}  {|c |c |c| c| c|}
 \hline 
 & Det. `$H$' & Det. `$V$' & Det. `$D$' & Det. `$A$' \\
  \hline
 Mode 1 & 0.08 & 0 & 0.00106 & 0.00106  \\
  \hline
 Mode 2 & 0 & 0.0008  & 0.0001 & 0.0001 \\
  \hline
 Mode 3 & 0.002 & 0.002 & 0.16 & 0 \\
  \hline
 Mode 4 & 0.002 & 0.002 & 0 & 0.04 \\
  \hline
\end{tabular}
\newline \\
\end{center}
\label{passive_mismatch_nopinhole} 
\end{table}

\begin{table}
\caption{Spatial-mode-dependent mismatch observed in Ref.~\cite{Shihan2015}, 
when there is a pinhole. Different columns are for different detectors labelled 
and shown in Fig.~\ref{setup}(b). Different rows are mismatched efficiencies 
for different spatial modes.}
\begin{center}
\begin{tabular}  {|c |c |c| c| c|}
 \hline 
 & Det. `$H$' & Det. `$V$' & Det. `$D$' & Det. `$A$' \\
  \hline
 Mode 1 & 0.0004 & 0 & 0.0002 & 0.0002  \\
  \hline
 Mode 2 & 0 & 0.0004  & 0.00033 & 0.00033 \\
  \hline
 Mode 3 & 0.0002 & 0.0002 & 0.0004 & 0 \\
  \hline
 Mode 4 & 0.00033 & 0.00033 & 0 & 0.0004 \\
  \hline
\end{tabular}
\newline \\
\end{center}
\label{passive_mismatch_pinhole} 
\end{table}

\section{Conclusion}
\label{conclusion}
Many methods for verifying entanglement assume that various detectors
involved in an experimental setup have the same efficiency and that the 
dimension of the quantum system is fixed and known. However, in practice, 
the efficiencies of detectors involved in a setup usually do not take the 
same value, and the tested optical system is not well characterized. 
To address these problems, one can apply device-independent criteria, 
such as violations of Bell inequalities. However, these device-independent 
methods are not robust against transmission loss. Hence, they cannot verify 
many entangled states that appear in practice.  Here, we present a method 
for verifying entanglement when the efficiency mismatch is characterized.  
Our method works without the knowledge of the system dimension. The method 
exploits relationships between measurement POVM elements, particularly 
those relationships between the projections of measurement POVM elements onto the  
subspace that contains only a few of photons. The projections contain 
efficiency-mismatch information, and their expectation values are connected 
with experimental observations by inequalities. Hence, our method can take 
advantage of efficiency-mismatch information to verify entanglement.  

We implement the method by exploiting the projections of measurement POVM elements 
onto the ($\leq 2$)-photons subspace. We expect that the entanglement-verification 
power of our method becomes stronger if a higher-photon-number subspace is 
considered. We demonstrate our method with simulations. The results show that 
our method can verify entanglement even if there exists spatial-mode-dependent 
mismatch, which could be induced by an adversary in the QKD scenario. Moreover,  
the results show that our method is robust against transmission loss, particularly  
when the projections of measurement POVM elements onto the two-photons subspace are exploited. 
For the no-mismatch case, there is another method for verifying entanglement based  
on squashing models~\cite{normand2008, Kiyoshi2008}. This method also does not 
require any system-dimension information. The simulation results show that our 
method can improve in some cases over squashing models for verifying entanglement 
(see Fig.~\ref{squash}). 

We have addressed the problem of verifying entanglement with efficiency mismatch
even without knowing the dimension of the system. Future work is required to adapt  
the method to prove the security of QKD protocols with efficiency mismatch. It is  
also desirable to have an analytical proof of the monotonic behaviours of the 
double-click, effective-error, or cross-click probabilities as functions 
of the number of photons arriving at Bob, as demonstrated in  
Figs.~\ref{active_doubleclick},~\ref{active_error}, and~\ref{passive_crossclick}. 
These monotonic behaviours are important for taking advantage of operators in  
the subspace that contains only a few of photons.

\begin{acknowledgments}
We thank Shihan Sajeed, Poompong Chaiwongkhot, Vadim Makarov, and Scott 
Glancy for useful discussions and comments.  We gratefully acknowledge supports
through the Office of Naval Research (ONR), the Ontario Research Fund (ORF), 
the Natural Sciences and Engineering Research Council of Canada (NSERC), and 
Industry Canada.
\end{acknowledgments}

\appendix
\section*{Appendix}
\label{sect:appendix}

\subsection{Operators exploited for constructing EVMs}
\label{sect:operators}
In the main text, we discuss only the operators exploited 
in the case of the active-detection scheme with one spatial 
mode. Here, we will study other cases considered in the paper. 

\subsubsection{Active-detection scheme with two spatial modes}
\label{active_operators_2spatial}

The idea behind constructing EVMs with two spatial modes (e.g., for the case
of the mismatch model in Table~\ref{active_mismatch}) is the same as that with one  
spatial mode studied in Sects.~\ref{active_operators_1} and \ref{active_operators_2} 
of the main text. We consider both measurement POVM elements and ideal operators 
in the $(\leq 2)$-photons subspace for constructing an EVM.  However, moving 
on to the two-spatial-modes case, measurement POVMs change their expressions. 
Also, more ideal operators in the $(\leq 2)$-photons subspace can be exploited. 
Let us discuss them in detail. 

First, considering the tensor-product structure over the two spatial modes, 
the expressions of measurement POVM elements change. For example, as long  
as there is a single click in one of the two spatial modes, it will contribute  
to the corresponding single-click event in experimental observations. So, the POVM 
elements for the single-click events in the $H/V$ basis are 
\begin{equation}
M_{SC}=M_{SC,1}\otimes M_{\emptyset,2}^{+}+ M_{\emptyset,1}^{+}\otimes M_{SC,2}+M_{SC,1}\otimes M_{SC,2},
\label{eq:active_2mode_POVMs}
\end{equation}
where the subscript `$SC$' can be either `$H$' or `$V$', and the subscripts `1' and `2'  
denote the two different spatial modes. For each spatial mode $i$, $i=1$, $2$, 
the expressions of $M_{H,i}, M_{V,i}$ and $M_{\emptyset,i}^{+}$ are the same as those 
in Eq.~\eqref{eq:active_HV_POVMs} with the replacement of detection efficiencies 
$\eta_{H/D}$ and $\eta_{V/A}$ by those efficiencies $\eta_{H/D,i}$ and $\eta_{V/A,i}$ 
for the spatial mode $i$. In a similar way, we can write down the other four POVM 
elements $M_D$, $M_A$, $M_{HV}$ and $M_{DA}$ required for constructing an EVM. 
Note that the measurement device cannot measure a photon in a basis that 
involves a superposition of different spatial modes. So, measurement POVM elements
such as those in Eq.~\eqref{eq:active_2mode_POVMs} are block-diagonal with respect 
to various photon-number subspaces where the number of photons in each spatial mode
is specified.

Second, as for the one-spatial-mode case, the projections of POVM elements onto the 
zero-photon, one-photon, or two-photons subspaces are linear combinations of ideal 
operators in these subspaces. The projection onto the zero-photon subspace 
is still expressed by the ideal operator in the set $\mathbf{\mathcal{S}}_{0}$ of  
Eq.~\eqref{eq:op_set0}. However, to express the projections onto the one-photon and 
two-photons subspaces, we need additional ideal operators. For the one-photon case, 
a photon can either lie in the spatial mode 1 or 2. So the set of ideal operators 
considered in Eq.~\eqref{eq:op_set1} expands to two parallel sets
\begin{align}
& \mathbf{\mathcal{S}}_{1,1}\equiv \{I_{2\times 2,1}, \tilde M_{H,1}^{(1)}, \tilde M_{D,1}^{(1)}, \sigma_{y,1}\}\otimes \Ket{\rm{Vac}}_2\Bra{\rm{Vac}} \label{eq:op_set1_mode1} 
\end{align}
and
\begin{align}
& \mathbf{\mathcal{S}}_{1,2}\equiv \Ket{\rm{Vac}}_1\Bra{\rm{Vac}} \otimes \{I_{2\times 2,2}, \tilde M_{H,2}^{(1)}, \tilde M_{D,2}^{(1)}, \sigma_{y,2}\}, \label{eq:op_set1_mode2}
\end{align}
where the second subscripts `1' and `2' denote the two spatial modes. 
Note that in Eqs.~\eqref{eq:op_set1_mode1} and~\eqref{eq:op_set1_mode2}, the ideal 
operators for each spatial mode have the same expressions as those for the one-spatial-mode 
case; the same is true for the ideal operators in the equations below. For the 
two-photons case, there are three possibilities: Both photons are in the same spatial 
mode 1 or 2, and one photon is in the spatial mode 1 and the other is in the spatial 
mode 2.  For each of the first two possibilities, the set of ideal operators as  
defined in Eq.~\eqref{eq:op_set2} becomes
\begin{align}
\mathbf{\mathcal{S}}_{2,1}\equiv &\{I_{3\times 3,1}, \tilde M_{H,1}^{(2)}, \tilde M_{V,1}^{(2)}, \tilde M_{D,1}^{(2)}, \tilde M_{A,1}^{(2)}, S_{y,1}\} \notag \\
& \otimes \Ket{\rm{Vac}}_2\Bra{\rm{Vac}} \label{eq:op_set2_mode1} 
\end{align}
or
\begin{align}
\mathbf{\mathcal{S}}_{2,2}\equiv& \Ket{\rm{Vac}}_1\Bra{\rm{Vac}}\otimes \notag \\
& \{I_{3\times 3,2}, \tilde M_{H,2}^{(2)}, \tilde M_{V,2}^{(2)}, \tilde M_{D,2}^{(2)},\tilde M_{A,2}^{(2)}, S_{y,2}\}. \label{eq:op_set2_mode2} 
\end{align}
For the case that each spatial mode holds one photon, each spatial mode is a qubit 
system. To describe the projections of measurement POVMs onto this subspace and  
also to study their operator relationships, we need to consider the following set 
of ideal operators: 
\begin{align}
&\mathbf{\mathcal{S}}_{2,1+2}\equiv \{I_{2\times 2,1}\otimes I_{2\times 2,2}, \tilde M_{H,1}^{(1)}\otimes I_{2\times 2,2}, \tilde M_{D,1}^{(1)} \otimes I_{2\times 2,2}, \notag \\
&\sigma_{y,1}\otimes I_{2\times 2,2}, I_{2\times 2,1}\otimes \tilde M_{H,2}^{(1)}, I_{2\times 2,1}\otimes \tilde M_{D,2}^{(1)}, I_{2\times 2,1}\otimes \sigma_{y,2}\}.
\label{eq:op_set2_mode12}
\end{align}

Moreover, in order to exploit the ideal operators within the sets $\mathbf{\mathcal{S}}_{0}$, 
$\mathbf{\mathcal{S}}_{1,1}$, $\mathbf{\mathcal{S}}_{1,2}$, $\mathbf{\mathcal{S}}_{2,1}$,  
$\mathbf{\mathcal{S}}_{2,2}$, and $\mathbf{\mathcal{S}}_{2,1+2}$, we need to know the 
relations between these ideal operators and measurement POVM elements. Once we know the  
relations, we can bound expectation values of these ideal operators based on 
experimental observations. As in the one-spatial-mode case, the idea is to express 
the projections of measurement POVM elements onto the ($\leq 2$)-photons subspace as 
linear combinations of these ideal operators. Considering the block-diagonal 
structure of measurement POVM elements with respect to various photon-number 
subspaces across the two spatial modes, then the relations are established. 
Therefore, to construct an EVM, we exploit both the measurement operators in  
the set $\mathbf{\mathcal{S}}$ of Eq.~\eqref{eq:op_set} and the ideal operators  
in the sets $\mathbf{\mathcal{S}}_{0}$ of Eq.~\eqref{eq:op_set0},  
$\mathbf{\mathcal{S}}_{1,1}$ of Eq.~\eqref{eq:op_set1_mode1}, 
$\mathbf{\mathcal{S}}_{1,2}$ of Eq.~\eqref{eq:op_set1_mode2}, 
$\mathbf{\mathcal{S}}_{2,1}$ of Eq.~\eqref{eq:op_set2_mode1},  
$\mathbf{\mathcal{S}}_{2,2}$ of Eq.~\eqref{eq:op_set2_mode2}, 
and $\mathbf{\mathcal{S}}_{2,1+2}$ of Eq.~\eqref{eq:op_set2_mode12}. 

Note that the constructed EVM can be restricted further by considering the 
relationships between operators in the above sets. The relationships between the  
measurement operators in the set $\mathbf{\mathcal{S}}$ are the same as those discussed 
in Sect.~\ref{active_operators_1} of the main text, even though the expressions of these 
operators change.  For each of the other operator sets, the relationships between 
operators therein can be derived from those between Pauli operators or between 
spin-1 operators (see Appendix~\ref{sect:spin_operators}). Moreover, any two operators 
from any two different sets of $\mathbf{\mathcal{S}}_{0}$, $\mathbf{\mathcal{S}}_{1,1}$, 
$\mathbf{\mathcal{S}}_{1,2}$,  $\mathbf{\mathcal{S}}_{2,1}$, $\mathbf{\mathcal{S}}_{2,2}$ 
and $\mathbf{\mathcal{S}}_{2,1+2}$ are orthogonal to each other.

\subsubsection{Passive-detection scheme}
\label{passive_operators}

In the passive-detection scheme as shown in Fig.~\ref{setup}(b), there are eight possible 
detection events in total: No click at any of the four detectors, click at only one of the  
four detectors (single click), clicks at the two detectors at the same output arm of the beam 
splitter (double click), and clicks at two or more detectors at different output arms of the  
beam splitter (cross click).  

Let us first consider the operators when there is only one spatial mode, since these operators 
are building blocks for more general cases. In contrast to the active-detection case, the explicit 
expressions of measurement POVM elements in the full state space are quite lengthy  
(see Appendix~\ref{sect:povm_with_mismatch}). Denote these POVM elements by $M_\emptyset$, 
$M_H$, $M_V$, $M_D$, $M_A$, $M_{HV}$, $M_{DA}$, and $M_{CC}$, where the subscripts indicate 
measurement outcomes and `CC' means cross click. These operators are positive-semidefinite 
and block-diagonal with respect to various photon-number subspaces. Furthermore, since all 
POVM elements have eigenvalues between 0 and 1, they satisfy the following relationships:
\begin{equation}
M_i\geq M_iM_i\geq 0, \label{eq:passive_rela}
\end{equation}
where $i$ can be $\emptyset$, $H$, $V$, $D$, $A$, $HV$, $DA$, or $CC$. 
Considering that 
$M_\emptyset+M_H+M_V+M_D+M_A+M_{HV}+M_{DA}+M_{CC}=I$ where $I$ is the identity operator in the 
full state space, we utilize the operator set
\begin{equation}
\mathbf{\mathcal{S}}\equiv\{I, M_H, M_V, M_{HV}, M_D, M_A, M_{DA}, M_{CC}\}, 
\label{eq:passive_op_set}
\end{equation}
to construct an EVM. 

We also consider the projections of the above POVM elements onto the ($\leq 2$)-photons subspace, 
and exploit relationships between these projections. As in the active-detection case, these 
projections are linear combinations of ideal operators in the same subspace, where the combination 
coefficients depend on mismatched efficiencies (see Appendix~\ref{sect:povm_with_mismatch} for 
more details). These ideal operators include those in the operator sets 
$\mathbf{\mathcal{S}}_{0}$, $\mathbf{\mathcal{S}}_{1}$ and $\mathbf{\mathcal{S}}_{2}$ in 
Eqs.~\eqref{eq:op_set0},~\eqref{eq:op_set1} and~\eqref{eq:op_set2}. As for the active-detection 
scheme, we can bound the expectation values of these ideal operators and exploit the relationships 
between them. Hence, in addition to the measurement operators in the set $\mathbf{\mathcal{S}}$ of 
Eq.~\eqref{eq:passive_op_set} we also utilize the ideal operators in the sets 
$\mathbf{\mathcal{S}}_{0}$, $\mathbf{\mathcal{S}}_{1}$ and $\mathbf{\mathcal{S}}_{2}$
(Eqs.~\eqref{eq:op_set0} to \eqref{eq:op_set2}) for constructing an EVM.

Next, we consider the four-spatial-modes case (e.g., the case of the mismatch model in 
Table~\ref{passive_mismatch}). Considering the tensor-product structure over the four spatial 
modes, the no-click POVM element is 
\begin{equation}
M_{\emptyset}=M_{\emptyset,1}\otimes M_{\emptyset,2}\otimes M_{\emptyset,3}\otimes M_{\emptyset,4},
\label{eq:no_click_povm}
\end{equation} 
where the subscripts `1', `2', `3' and `4' are the indices of different spatial modes. 
Moreover, as long as there is a single click in one of the four spatial modes, it will contribute to the 
corresponding single-click event in experimental observations. So the single-click POVM 
elements are 
\begin{equation}
M_{SC}= M_{?,1}\otimes M_{?,2}\otimes M_{?,3}\otimes M_{?,4},
\label{eq:single_click_povm}
\end{equation} 
where the subscript `$SC$' can be `$H$', `$V$', `$D$' or `$A$', and the notation `$?$' can be 
either single click `$SC$' or no detection `$\emptyset$' but at least one of $M_{?,1}$, $M_{?,2}$, 
$M_{?,3}$ and $M_{?,4}$ must be the single-click POVM element for the one-spatial-mode case. 
In a similar way, we can write down the expressions of $M_{HV}$, $M_{DA}$ and $M_{CC}$. 
Although measurement POVM elements become more complicated as compared with the  
one-spatial-mode case, they still satisfy the relationships in Eq.~\eqref{eq:passive_rela}. 

To construct an EVM, we also consider the projections of measurement POVM elements onto the 
zero-photon and one-photon subspaces. (Unlike the one-spatial-mode case, 
we do not consider the projections onto the two-photons subspace due to the complexity of both 
these operators and their relationships.) To express these projections, we need ideal 
operators in the zero-photon and one-photon subspaces. For the zero-photon case, 
the ideal operator needed is in the set $\mathbf{\mathcal{S}}_{0}$ of Eq.~\eqref{eq:op_set0}. 
For the one-photon case, since the photon can be in any one of the four spatial modes, we need 
the following sets of ideal operators:
\begin{align}
\mathbf{\mathcal{S}}_{1,1}\equiv &\{I_{2\times 2,1}, \tilde M_{H,1}^{(1)}, \tilde M_{D,1}^{(1)}, \sigma_{y,1}\}\otimes \Ket{\rm{Vac}}_2\Bra{\rm{Vac}} \notag \notag \\
& \otimes \Ket{\rm{Vac}}_3\Bra{\rm{Vac}}\otimes \Ket{\rm{Vac}}_4\Bra{\rm{Vac}}, \label{eq:passive_op_set1_mode1} \\
\mathbf{\mathcal{S}}_{1,2}\equiv & \Ket{\rm{Vac}}_1\Bra{\rm{Vac}} \otimes \{I_{2\times 2,2}, \tilde M_{H,2}^{(1)}, \tilde M_{D,2}^{(1)}, \sigma_{y,2}\} \notag \\
& \otimes\Ket{\rm{Vac}}_3\Bra{\rm{Vac}}\otimes \Ket{\rm{Vac}}_4\Bra{\rm{Vac}}, \label{eq:passive_op_set1_mode2} \\
\mathbf{\mathcal{S}}_{1,3}\equiv & \Ket{\rm{Vac}}_1\Bra{\rm{Vac}} \otimes \Ket{\rm{Vac}}_2\Bra{\rm{Vac}} \notag \\
& \otimes \{I_{2\times 2,3}, \tilde M_{H,3}^{(1)}, \tilde M_{D,3}^{(1)}, \sigma_{y,3}\} \otimes \Ket{\rm{Vac}}_4\Bra{\rm{Vac}}, \label{eq:passive_op_set1_mode3} 
\end{align}
and
\begin{align}
\mathbf{\mathcal{S}}_{1,4}\equiv & \Ket{\rm{Vac}}_1\Bra{\rm{Vac}} \otimes \Ket{\rm{Vac}}_2\Bra{\rm{Vac}} \notag \\
& \otimes \Ket{\rm{Vac}}_3\Bra{\rm{Vac}} \otimes \{I_{2\times 2,4}, \tilde M_{H,4}^{(1)}, \tilde M_{D,4}^{(1)}, \sigma_{y,4}\}. \label{eq:passive_op_set1_mode4} 
\end{align}
The relationships between operators in each of the above sets can be derived from those 
between Pauli operators (see Appendix~\ref{sect:spin_operators}), while any two operators 
from any two different sets of $\mathbf{\mathcal{S}}_{0}$, $\mathbf{\mathcal{S}}_{1,1}$, 
$\mathbf{\mathcal{S}}_{1,2}$, $\mathbf{\mathcal{S}}_{1,3}$ and $\mathbf{\mathcal{S}}_{1,4}$
are orthogonal to each other. Moreover, as for the active-detection scheme, we 
can bound the expectation values of these ideal operators based on experimental observations. 
Therefore, to construct an EVM for the four-spatial-modes case, we utilize not only the 
measurement operators in the set $\mathbf{\mathcal{S}}$ of Eq.~\eqref{eq:passive_op_set} but 
also the ideal operators in the sets $\mathbf{\mathcal{S}}_{0}$ of Eq.~\eqref{eq:op_set0}, 
$\mathbf{\mathcal{S}}_{1,1}$ of Eq.~\eqref{eq:passive_op_set1_mode1}, 
$\mathbf{\mathcal{S}}_{1,2}$ of Eq.~\eqref{eq:passive_op_set1_mode2},
$\mathbf{\mathcal{S}}_{1,3}$ of Eq.~\eqref{eq:passive_op_set1_mode3} and 
$\mathbf{\mathcal{S}}_{1,4}$ of Eq.~\eqref{eq:passive_op_set1_mode4}.

\subsection{POVM elements with mismatched efficiencies}
\label{sect:povm_with_mismatch}

\begin{figure}[htb!]
   \includegraphics[scale=0.58,viewport=10.5cm 13.5cm 16cm 22.5cm]{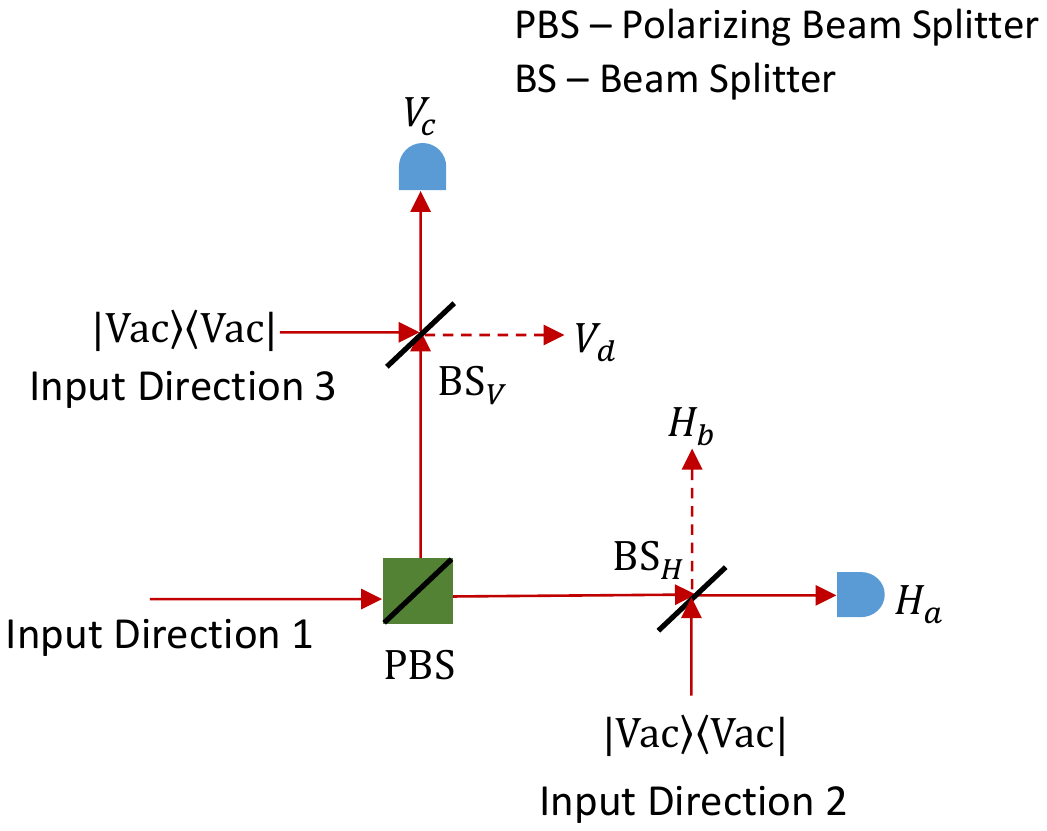} 
   \caption{Description of the measurement in the $H/V$ basis under
   the active-detection scheme. Suppose that the detection efficiencies 
   associated with the outcomes $H$ and $V$ are $\eta_{H/D}$ and $\eta_{V/A}$, 
   respectively, and that the two detectors shown are ideal. Then, the 
   beam splitters BS$_H$ and BS$_V$ have transmission coefficients
   $\sqrt{\eta_{H/D}}$ and $\sqrt{\eta_{V/A}}$, respectively. } 
   \label{povm_active}
\end{figure}

We first consider the active-detection scheme as shown in Fig.~\ref{setup}(a). 
Suppose that the two threshold detectors have efficiencies $\eta_{H/D}$ and 
$\eta_{V/A}$, respectively. Recall that a real threshold detector can be described 
by an ideal threshold detector with a beam splitter in front whose transmission 
coefficient is equal to the square root of the real detector's efficiency~\cite{Fox}. 
It then turns out that the measurement in the $H/V$ basis can be described by an optical 
device with three input spatial directions labelled by numbers and four output 
spatial directions labelled by lower-case letters, as shown in Fig.~\ref{povm_active}.  
In order to realize the measurement, however, only the input spatial direction 1  
has incoming optical signals, whereas no signals travel along the other two input 
spatial directions 2 and 3.  To get an outcome, the output modes $H_a$ and $V_c$ 
are measured with ideal threshold detectors, whereas the other two output modes 
$H_b$ and $V_d$ are not detected corresponding to the loss in the measurement 
process.  (Note that here and later we use both the polarization degree of 
freedom and the spatial degree of freedom to label an input or output mode.) 
Therefore, it is straightforward to write down POVM elements in the output-modes 
basis $\{\Ket{n_{H_a}, n_{H_b}, n_{V_c}, n_{V_d}}, n_{H_a}, n_{H_b}, n_{V_c}, 
n_{V_d}=0,1,2,...\}$. For example, the POVM element for the single-click event 
`$H$' is written as 
\begin{align}
M_H=&\sum_{n_{H_a}=1}^{\infty}\sum_{n_{H_b}=0}^{\infty}\sum_{n_{V_d}=0}^{\infty} \notag \\ 
&\Ket{n_{H_a},n_{H_b},0_{V_c}, n_{V_d}}\Bra{n_{H_a},n_{H_b},0_{V_c}, n_{V_d}}.
\label{povm_active_h_output1}
\end{align}
Using the fact $\Ket{n}=\frac{1}{\sqrt{n!}} (\hat a^{\dagger})^n \Ket{0}$, 
Eq.~\eqref{povm_active_h_output1} becomes  
\begin{align}
M_H=&\sum_{n_{H_a}=1}^{\infty}\sum_{n_{H_b}=0}^{\infty}\sum_{n_{V_d}=0}^{\infty} \frac{1}{n_{H_a}!\,n_{H_b}!\,n_{V_d}!} \notag \\
&\left(\hat a_{H_a}^{\dagger}\right)^{n_{H_a}} \left(\hat a_{H_b}^{\dagger}\right)^{n_{H_b}} \left(\hat a_{V_d}^{\dagger}\right)^{n_{V_d}}  \Ket{0}\Bra{0} \notag \\ 
& \left(\hat a_{H_a}\right)^{n_{H_a}}\left(\hat a_{H_b}\right)^{n_{H_b}}\left(\hat a_{V_d}\right)^{n_{V_d}},
\label{povm_active_h_output2}
\end{align}
where $\hat a $ and $\hat a^\dagger$ are annihilation and creation operators 
associated with an optical mode, respectively.

To express the POVM element $M_H$ in the input-modes basis, we utilize the relations 
between the creation operators of the input and output modes of a beam splitter. Specifically, 
the relations
\begin{align}
& \hat a_{H_a}^{\dagger}=\sqrt{\eta_{H/D}} \hat a_{H_1}^{\dagger}+\sqrt{1-\eta_{H/D}} \hat a_{H_2}^{\dagger}, \notag \\
& \hat a_{H_b}^{\dagger}=-\sqrt{1-\eta_{H/D}} \hat a_{H_1}^{\dagger}+\sqrt{\eta_{H/D}} \hat a_{H_2}^{\dagger}, \text{ and} \notag \\
& \hat a_{V_d}^{\dagger}=-\sqrt{1-\eta_{V/A}} \hat a_{V_1}^{\dagger}+\sqrt{\eta_{V/A}} \hat a_{V_3}^{\dagger}.
\label{BS_relation}
\end{align}
Using Eq.~\eqref{BS_relation}, we can express $M_H$ in terms of the creation operators 
$\hat a_{H_1}^{\dagger}, \hat a_{V_1}^{\dagger}$, $\hat a_{H_2}^{\dagger}$ and 
$\hat a_{V_3}^{\dagger}$ and the annihilation operators $\hat a_{H_1}, \hat a_{V_1}, 
\hat a_{H_2}$ and $\hat a_{V_3}$. Further, to fulfill the physical condition that there 
is no incoming optical signal travelling along the spatial directions 2 or 3, we need to 
pick only the terms where there is no appearance of any of the operators $\hat a_{H_2}^{\dagger}$, 
$\hat a_{V_3}^{\dagger}$, $\hat a_{H_2}$, and $\hat a_{V_3}$. Therefore, we get that 
\begin{align}
M_H=&\sum_{n_{H_a}=1}^{\infty}\sum_{n_{H_b}=0}^{\infty}\sum_{n_{V_d}=0}^{\infty} \frac{1}{n_{H_a}!\,n_{H_b}!\,n_{V_d}!} \notag \\
& (\eta_{H/D})^{n_{H_a}}(1-\eta_{H/D})^{n_{H_b}} (1-\eta_{V/A})^{n_{V_d}} \left( \hat a_{H_1}^{\dagger}\right)^{n_{H_a}+n_{H_b}}  \notag \\
& \left(\hat a_{V_1}^{\dagger}\right)^{n_{V_d}} \Ket{0}\Bra{0}  \left( \hat a_{H_1}\right)^{n_{H_a}+n_{H_b}}\left(\hat a_{V_1}\right)^{n_{V_d}}.
\label{povm_active_h_input1}
\end{align}
Using the fact that $(\hat a^{\dagger})^n \Ket{0}=\sqrt{n!}\Ket{n}$, and setting that 
$n_{H_1}=n_{H_a}+n_{H_b}$ and $n_{V_1}=n_{V_d}$, we can simplify the above equation to   
\begin{align}
M_H=&\sum_{n_{H_1}=1}^{\infty}\sum_{n_{V_1}=0}^{\infty}\left(1-(1-\eta_{H/D})^{n_{H_1}}\right)(1-\eta_{V/A})^{n_{V_1}} \notag \\ 
&\Ket{n_{H_1},n_{V_1}}\Bra{n_{H_1},n_{V_1}}.
\label{povm_active_h_input2}
\end{align}
The above equation is the same as that in Eq.~\eqref{eq:active_HV_POVMs} of the main text, 
except that each polarization mode has a subscript `1' indicating the input spatial direction. 

Likewise, we can derive the other POVM elements for the measurement in the $H/V$ basis, and 
the results are as follows:
\begin{align}
M_V=&\sum_{n_{H_1}=0}^{\infty}\sum_{n_{V_1}=1}^{\infty}(1-\eta_{H/D})^{n_{H_1}}\left(1-(1-\eta_{V/A})^{n_{V_1}}\right) \notag \\
&\Ket{n_{H_1},n_{V_1}}\Bra{n_{H_1},n_{V_1}}, \notag \\
M_{HV}=&\sum_{n_{H_1}=1}^{\infty}\sum_{n_{V_1}=1}^{\infty}\left(1-(1-\eta_{H/D})^{n_{H_1}}\right)\left(1-(1-\eta_{V/A})^{n_{V_1}}\right)\notag \\
&\Ket{n_{H_1},n_{V_1}}\Bra{n_{H_1},n_{V_1}}, \text{ and}  \notag \\
M_{\emptyset}^{+}=& \sum_{n_{H_1}=0}^{\infty}\sum_{n_{V_1}=0}^{\infty}(1-\eta_{H/D})^{n_{H_1}}(1-\eta_{V/A})^{n_{V_1}} \notag \\
&\Ket{n_{H_1},n_{V_1}}\Bra{n_{H_1},n_{V_1}},
\label{eq:povm_active_hv_input}
\end{align}
where the superscript `$+$' denotes the $H/V$ basis.

\begin{figure}[htb!]
   \includegraphics[scale=0.58,viewport=8.5cm 12cm 14cm 22.5cm]{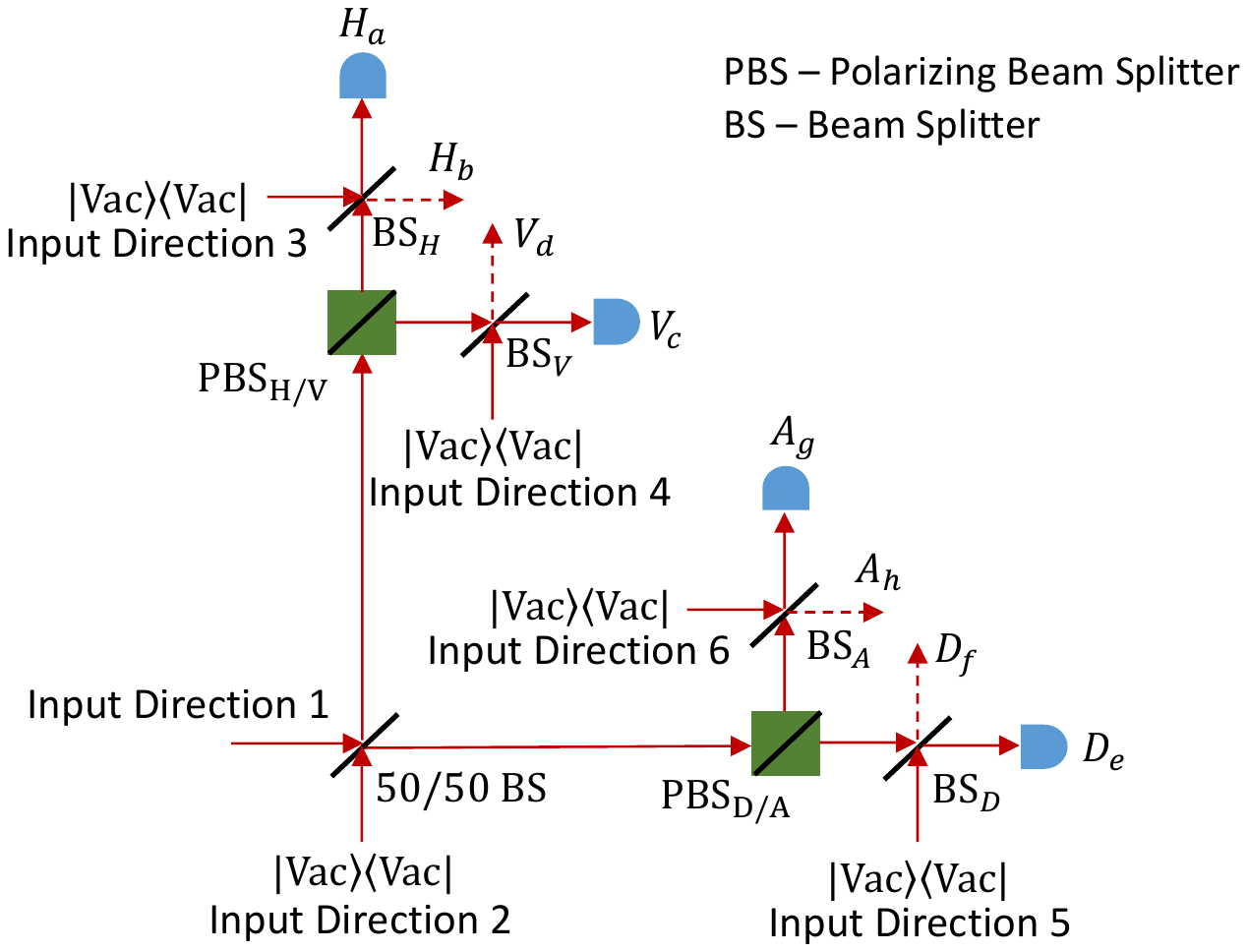} 
   \caption{Description of the measurement under the passive-detection 
   scheme. Suppose that the detection efficiencies associated with the 
   outcomes $H$, $V$, $D$ and $A$ are $\eta_H$, $\eta_V$, $\eta_D$ and $\eta_A$
   respectively, and that the four detectors shown are ideal. Then, the 
   beam splitters BS$_H$, BS$_V$, BS$_D$ and BS$_A$ have transmission coefficients
   $\sqrt{\eta_H}$, $\sqrt{\eta_V}$, $\sqrt{\eta_D}$ and $\sqrt{\eta_A}$, respectively.}  
   \label{povm_passive}
\end{figure}

We can follow the same procedure as above to derive the POVM elements 
for the measurement in the $D/A$ basis under the active-detection scheme. 
We skip the details and just list the POVM elements as follows:
\begin{align}
M_D=&\sum_{n_{D_1}=1}^{\infty}\sum_{n_{A_1}=0}^{\infty}\left(1-(1-\eta_{H/D})^{n_{D_1}}\right)(1-\eta_{V/A})^{n_{A_1}} \notag \\ 
&\Ket{n_{D_1},n_{A_1}}\Bra{n_{D_1},n_{A_1}}, \notag 
\end{align}
\begin{align}
M_A=&\sum_{n_{D_1}=0}^{\infty}\sum_{n_{A_1}=1}^{\infty}(1-\eta_{H/D})^{n_{D_1}}\left(1-(1-\eta_{V/A})^{n_{A_1}}\right) \notag \\
&\Ket{n_{D_1},n_{A_1}}\Bra{n_{D_1},n_{A_1}}, \notag \\
M_{DA}=&\sum_{n_{D_1}=1}^{\infty}\sum_{n_{A_1}=1}^{\infty}\left(1-(1-\eta_{H/D})^{n_{D_1}}\right)\left(1-(1-\eta_{V/A})^{n_{A_1}}\right)\notag \\
&\Ket{n_{D_1},n_{A_1}}\Bra{n_{D_1},n_{A_1}}, \text{ and} \notag \\
M_{\emptyset}^{\times}=& \sum_{n_{D_1}=0}^{\infty}\sum_{n_{A_1}=0}^{\infty}(1-\eta_{H/D})^{n_{D_1}}(1-\eta_{V/A})^{n_{A_1}} \notag \\
&\Ket{n_{D_1},n_{A_1}}\Bra{n_{D_1},n_{A_1}}.
\label{eq:povm_active_da_input}
\end{align}
Here, the superscript `$\times$' denotes the $D/A$ basis, and the subscript `1' 
of the polarization mode denotes the input spatial direction.

Next, let us study the passive-detection case. With the help of the beam-splitter 
model for a real detector with imperfect efficiency, we can see that the whole 
measurement can be described by an optical device with six input spatial 
directions and eight output spatial directions, as shown in Fig.~\ref{povm_passive}. 
To realize the measurement, only the input spatial direction 1 has incoming optical 
signals, and only the output modes $H_a$, $V_c$, $D_e$ and $A_g$ are measured (by 
ideal detectors). It is straightforward to see that in the output-modes basis 
$\{\Ket{n_{H_a},n_{H_b},n_{V_c}, n_{V_d}, n_{D_e},n_{D_f},n_{A_g}, n_{A_h}}, 
n_{H_a}, n_{H_b},n_{V_c}, \\
n_{V_d},n_{D_e}, n_{D_f}, n_{A_g}, n_{A_h}=0,1,2,...\}$,  
the POVM element for the single-click event `$H$' is written as
\begin{align}
M_H=&\sum_{n_{H_a}=1}^{\infty}\sum_{n_{H_b}=0}^{\infty}\sum_{n_{V_d}=0}^{\infty}\sum_{n_{D_f}=0}^{\infty} \sum_{n_{A_h}=0}^{\infty} \notag \\ 
&\Ket{n_{H_a},n_{H_b},0_{V_c}, n_{V_d}, 0_{D_e},n_{D_f},0_{A_g}, n_{A_h}} \notag \\
& \Bra{n_{H_a},n_{H_b},0_{V_c}, n_{V_d}, 0_{D_e},n_{D_f},0_{A_g}, n_{A_h}}.
\label{povm_passive_h_output}
\end{align}

To express the above POVM element $M_H$ in the input-modes basis, as for the 
active-detection case, we use the fact that 
$\Ket{n}=\frac{1}{\sqrt{n!}}(\hat a^{\dagger})^n \Ket{0}$ and the relations 
between the creation operators of the input and output modes of a beam splitter. 
Here, we skip the lengthy but not difficult details, and just write down the 
final expression of $M_H$ as follows: 
\begin{widetext}
\begin{align}
M_H=&\sum_{n_{H_a}=1}^{\infty}\sum_{n_{H_b}=0}^{\infty}\sum_{n_{V_d}=0}^{\infty}\sum_{n_{D_f}=0}^{\infty} \sum_{n_{A_h}=0}^{\infty} \left(\frac{1}{2}\right)^{n_{H_a}+n_{H_b}+n_{V_d}+n_{D_f}+n_{A_h}} \notag \\
&\frac{1}{n_{H_a}!\,n_{H_b}!\,n_{V_d}!\,n_{D_f}!\,n_{A_h}!} (\eta_H)^{n_{H_a}} (1-\eta_H)^{n_{H_b}} (1-\eta_V)^{n_{V_d}} (1-\eta_D)^{n_{D_f}} (1-\eta_A)^{n_{A_h}}  \notag \\
& \left( \hat a_{H_1}^{\dagger}\right)^{n_{H_a}+n_{H_b}} \left(\hat a_{V_1}^{\dagger}\right)^{n_{V_d}} \left(\hat a_{D_1}^{\dagger}\right)^{n_{D_f}} \left(\hat a_{A_1}^{\dagger}\right)^{n_{A_h}}  \Ket{0}\Bra{0} \left( \hat a_{H_1}\right)^{n_{H_a}+n_{H_b}}\left(\hat a_{V_1}\right)^{n_{V_d}} \left(\hat a_{D_1}\right)^{n_{D_f}} \left(\hat a_{A_1}\right)^{n_{A_h}}.
\label{povm_passive_h_input}
\end{align}

Likewise, we can get the following results:
\begin{align}
M_V=&\sum_{n_{H_b}=0}^{\infty}\sum_{n_{V_c}=1}^{\infty}\sum_{n_{V_d}=0}^{\infty}\sum_{n_{D_f}=0}^{\infty} \sum_{n_{A_h}=0}^{\infty} 
\left(\frac{1}{2}\right)^{n_{H_b}+n_{V_c}+n_{V_d}+n_{D_f}+n_{A_h}} \notag \\
&\frac{1}{n_{H_b}!\,n_{V_c}!\,n_{V_d}!\,n_{D_f}!\,n_{A_h}!} (1-\eta_H)^{n_{H_b}} (\eta_V)^{n_{V_c}} 
(1-\eta_V)^{n_{V_d}} (1-\eta_D)^{n_{D_f}} (1-\eta_A)^{n_{A_h}}  \notag \\
& \left( \hat a_{H_1}^{\dagger}\right)^{n_{H_b}} \left(\hat a_{V_1}^{\dagger}\right)^{n_{V_c}+n_{V_d}} \left(\hat a_{D_1}^{\dagger}\right)^{n_{D_f}} \left(\hat a_{A_1}^{\dagger}\right)^{n_{A_h}}  
 \Ket{0}\Bra{0} \left( \hat a_{H_1}\right)^{n_{H_b}}\left(\hat a_{V_1}\right)^{n_{V_c}+n_{V_d}} \left(\hat a_{D_1}\right)^{n_{D_f}} \left(\hat a_{A_1}\right)^{n_{A_h}}, \notag \\
M_D=&\sum_{n_{H_b}=0}^{\infty}\sum_{n_{V_d}=0}^{\infty}\sum_{n_{D_e}=1}^{\infty}\sum_{n_{D_f}=0}^{\infty} \sum_{n_{A_h}=0}^{\infty} 
\left(\frac{1}{2}\right)^{n_{H_b}+n_{V_d}+n_{D_e}+n_{D_f}+n_{A_h}} \notag \\
&\frac{1}{n_{H_b}!\,n_{V_d}!\,n_{D_e}!\,n_{D_f}!\,n_{A_h}!} (1-\eta_H)^{n_{H_b}}(1-\eta_V)^{n_{V_d}} 
(\eta_D)^{n_{D_e}} (1-\eta_D)^{n_{D_f}} (1-\eta_A)^{n_{A_h}}  \notag \\
& \left( \hat a_{H_1}^{\dagger}\right)^{n_{H_b}} \left(\hat a_{V_1}^{\dagger}\right)^{n_{V_d}} \left(\hat a_{D_1}^{\dagger}\right)^{n_{D_e}+n_{D_f}} \left(\hat a_{A_1}^{\dagger}\right)^{n_{A_h}}  
\Ket{0}\Bra{0} \left( \hat a_{H_1}\right)^{n_{H_b}}\left(\hat a_{V_1}\right)^{n_{V_d}} \left(\hat a_{D_1}\right)^{n_{D_e}+n_{D_f}} \left(\hat a_{A_1}\right)^{n_{A_h}}, \notag\\
M_A=&\sum_{n_{H_b}=0}^{\infty}\sum_{n_{V_d}=0}^{\infty}\sum_{n_{D_f}=0}^{\infty} \sum_{n_{A_g}=1}^{\infty} \sum_{n_{A_h}=0}^{\infty} 
\left(\frac{1}{2}\right)^{n_{H_b}+n_{V_d}+n_{D_f}+n_{A_g}+n_{A_h}} \notag \\
&\frac{1}{n_{H_b}!\,n_{V_d}!\,n_{D_f}!\,n_{A_g}!\,n_{A_h}!} (1-\eta_H)^{n_{H_b}}(1-\eta_V)^{n_{V_d}} 
(1-\eta_D)^{n_{D_f}}  (\eta_A)^{n_{A_g}} (1-\eta_A)^{n_{A_h}}  \notag \\
& \left( \hat a_{H_1}^{\dagger}\right)^{n_{H_b}} \left(\hat a_{V_1}^{\dagger}\right)^{n_{V_d}} \left(\hat a_{D_1}^{\dagger}\right)^{n_{D_f}} \left(\hat a_{A_1}^{\dagger}\right)^{n_{A_g}+n_{A_h}}  
\Ket{0}\Bra{0} \left( \hat a_{H_1}\right)^{n_{H_b}}\left(\hat a_{V_1}\right)^{n_{V_d}} \left(\hat a_{D_1}\right)^{n_{D_f}} \left(\hat a_{A_1}\right)^{n_{A_g}+n_{A_h}}, \notag \\
M_{HV}=&\sum_{n_{H_a}=1}^{\infty}\sum_{n_{H_b}=0}^{\infty}\sum_{n_{V_c}=1}^{\infty}\sum_{n_{V_d}=0}^{\infty}\sum_{n_{D_f}=0}^{\infty} \sum_{n_{A_h}=0}^{\infty} 
\left(\frac{1}{2}\right)^{n_{H_a}+n_{H_b}+n_{V_c}+n_{V_d}+n_{D_f}+n_{A_h}} \notag \\
&\frac{1}{n_{H_a}!\,n_{H_b}!\,n_{V_c}!\,n_{V_d}!\,n_{D_f}!\,n_{A_h}!} (\eta_H)^{n_{H_a}}(1-\eta_H)^{n_{H_b}}  
(\eta_V)^{n_{V_c}} (1-\eta_V)^{n_{V_d}} (1-\eta_D)^{n_{D_f}} (1-\eta_A)^{n_{A_h}}  \notag \\
& \left( \hat a_{H_1}^{\dagger}\right)^{n_{H_a}+n_{H_b}} \left(\hat a_{V_1}^{\dagger}\right)^{n_{V_c}+n_{V_d}} \left(\hat a_{D_1}^{\dagger}\right)^{n_{D_f}} \left(\hat a_{A_1}^{\dagger}\right)^{n_{A_h}}  
\Ket{0}\Bra{0} \left( \hat a_{H_1}\right)^{n_{H_a}+n_{H_b}}\left(\hat a_{V_1}\right)^{n_{V_c}+n_{V_d}} \left(\hat a_{D_1}\right)^{n_{D_f}} \left(\hat a_{A_1}\right)^{n_{A_h}}, \notag
\end{align}
\end{widetext}

\begin{widetext}
\begin{align}
M_{DA}=&\sum_{n_{H_b}=0}^{\infty}\sum_{n_{V_d}=0}^{\infty}\sum_{n_{D_e}=1}^{\infty}\sum_{n_{D_f}=0}^{\infty} \sum_{n_{A_g}=1}^{\infty} \sum_{n_{A_h}=0}^{\infty} 
\left(\frac{1}{2}\right)^{n_{H_b}+n_{V_d}+n_{D_e}+n_{D_f}+n_{A_g}+n_{A_h}} \notag \\
&\frac{1}{n_{H_b}!\,n_{V_d}!\,n_{D_e}!\,n_{D_f}!\,n_{A_g}!\,n_{A_h}!} (1-\eta_H)^{n_{H_b}}(1-\eta_V)^{n_{V_d}} 
(\eta_D)^{n_{D_e}} (1-\eta_D)^{n_{D_f}} (\eta_A)^{n_{A_g}}  (1-\eta_A)^{n_{A_h}}  \notag \\
& \left( \hat a_{H_1}^{\dagger}\right)^{n_{H_b}} \left(\hat a_{V_1}^{\dagger}\right)^{n_{V_d}} \left(\hat a_{D_1}^{\dagger}\right)^{n_{D_e}+n_{D_f}} \left(\hat a_{A_1}^{\dagger}\right)^{n_{A_g}+n_{A_h}}  
\Ket{0}\Bra{0} \left( \hat a_{H_1}\right)^{n_{H_b}}\left(\hat a_{V_1}\right)^{n_{V_d}} \left(\hat a_{D_1}\right)^{n_{D_e}+n_{D_f}} \left(\hat a_{A_1}\right)^{n_{A_g}+n_{A_h}},  \notag \\
\text{and}  & \notag \\
M_{\emptyset}=&\sum_{n_{H_b}=0}^{\infty}\sum_{n_{V_d}=0}^{\infty}\sum_{n_{D_f}=0}^{\infty} \sum_{n_{A_h}=0}^{\infty} 
\left(\frac{1}{2}\right)^{n_{H_b}+n_{V_d}+n_{D_f}+n_{A_h}} \notag \\
&\frac{1}{n_{H_b}!\,n_{V_d}!\,n_{D_f}!\,n_{A_h}!} (1-\eta_H)^{n_{H_b}}  (1-\eta_V)^{n_{V_d}} 
(1-\eta_D)^{n_{D_f}} (1-\eta_A)^{n_{A_h}} \left( \hat a_{H_1}^{\dagger}\right)^{n_{H_b}} \left(\hat a_{V_1}^{\dagger}\right)^{n_{V_d}}  \notag \\
&  \left(\hat a_{D_1}^{\dagger}\right)^{n_{D_f}} \left(\hat a_{A_1}^{\dagger}\right)^{n_{A_h}}  \Ket{0}\Bra{0} \left( \hat a_{H_1}\right)^{n_{H_b}} 
\left(\hat a_{V_1}\right)^{n_{V_d}} \left(\hat a_{D_1}\right)^{n_{D_f}} \left(\hat a_{A_1}\right)^{n_{A_h}}.
\label{povm_passive_input}
\end{align}
\end{widetext}

There is no obvious simplification of Eqs.~\eqref{povm_passive_h_input} 
and~\eqref{povm_passive_input}, but from these expressions it is straightforward to
see that the POVM elements are linear combinations of ideal operators where the combination
coefficients are determined by the mismatched efficiencies $\eta_H$, $\eta_V$, $\eta_D$ and 
$\eta_A$. Note that we did not write down the POVM element $M_{CC}$ for cross-click events, 
since it can be inferred from the fact 
$M_\emptyset+M_H+M_V+M_D+M_A+M_{HV}+M_{DA}+M_{CC}=I$ where $I$ is the identity operator in 
the full state space.

\subsection{Relationships between ideal operators} 
\label{sect:spin_operators}

To figure out the relationships between the ideal operators in Eq.~\eqref{eq:active_12photon}, 
we note that these ideal operators are related to Pauli operators (or spin-$\frac{1}{2}$ operators)  
\begin{widetext}
\begin{equation}
 \sigma_x=\left(\begin{array}{c c}
       0 &   1     \\
       1 &   0     
     \end{array}\right), 
 \sigma_y=\left(\begin{array}{c c}
       0 &   -i     \\
       i &   0     
     \end{array} \right), \text{ and }    
\sigma_z=\left(\begin{array}{c c}
       1 &   0     \\
       0 &   -1     
     \end{array} \right),
     \label{eq:pauli_xz} 
\end{equation}
\end{widetext}
and spin-1 operators 
\begin{widetext}
\begin{equation}
S_x=\frac{1}{\sqrt{2}}\left(\begin{array}{c c c}
       0 &   1  & 0   \\
       1 &   0  & 1   \\
       0 &   1  & 0  
     \end{array}\right),
S_y=\frac{1}{\sqrt{2}}\left(\begin{array}{c c c}
       0 &   -i  & 0   \\
       i &   0  & -i   \\
       0 &   i  & 0  
     \end{array}\right), \text{ and }
S_z=\left(\begin{array}{c c c}
       1 &   0  & 0   \\
       0 &   0  & 0   \\
       0 &   0  & -1  
     \end{array}\right).
     \label{eq:spin_xz}
\end{equation}
\end{widetext}
Particularly, in the basis $\{\Ket{1_H,0_V}, \Ket{0_H,1_V}\}$ of the one-photon 
subspace we have
\begin{equation}
2\tilde M_H^{(1)}-I_{2\times 2}=\sigma_z \text{ and } 2\tilde M_D^{(1)}-I_{2\times 2}=\sigma_x,
\label{1-photon_rela}
\end{equation}
and in the basis $\{\Ket{2_H,0_V}, \Ket{1_H,1_V}, \Ket{0_H,2_V}\}$ of the two-photons 
subspace we have 
\begin{equation}
\tilde M_H^{(2)}-\tilde M_V^{(2)}=S_z \text{ and } \tilde M_D^{(2)}-\tilde M_A^{(2)}=S_x. 
\label{2-photons_rela}
\end{equation}
So, we can derive the relationships between the ideal operators in 
Eq.~\eqref{eq:active_12photon} from those between Pauli operators or between spin-1 
operators, for example, from
\begin{equation}
\sigma_a\sigma_b=i\varepsilon_{abc}\sigma_c+\delta_{ab}I_{2\times 2}, \text{ or } [S_a,S_b]=i\varepsilon_{abc}S_c,
\label{eq:active_commutation}
\end{equation} 
where $\varepsilon_{abc}$ is the Levi-Civita symbol, $\delta_{ab}$ is the Kronecker delta 
function, and each of $a$, $b$ and $c$ can be $x$, $y$ or $z$.

In the same way, we can derive the relationships between the ideal operators in 
Eqs.~\eqref{eq:op_set1_mode1} to \eqref{eq:op_set2_mode12} and Eqs.~\eqref{eq:passive_op_set1_mode1}
to \eqref{eq:passive_op_set1_mode4}.


\subsection{Proof of the equivalence of the descriptions in Fig.~\ref{active_equivalent}}
\label{sect:proof_active_equiva}

\begin{figure}[htb!]
   \includegraphics[scale=0.58,viewport=3cm 13.5cm 18cm 20cm]{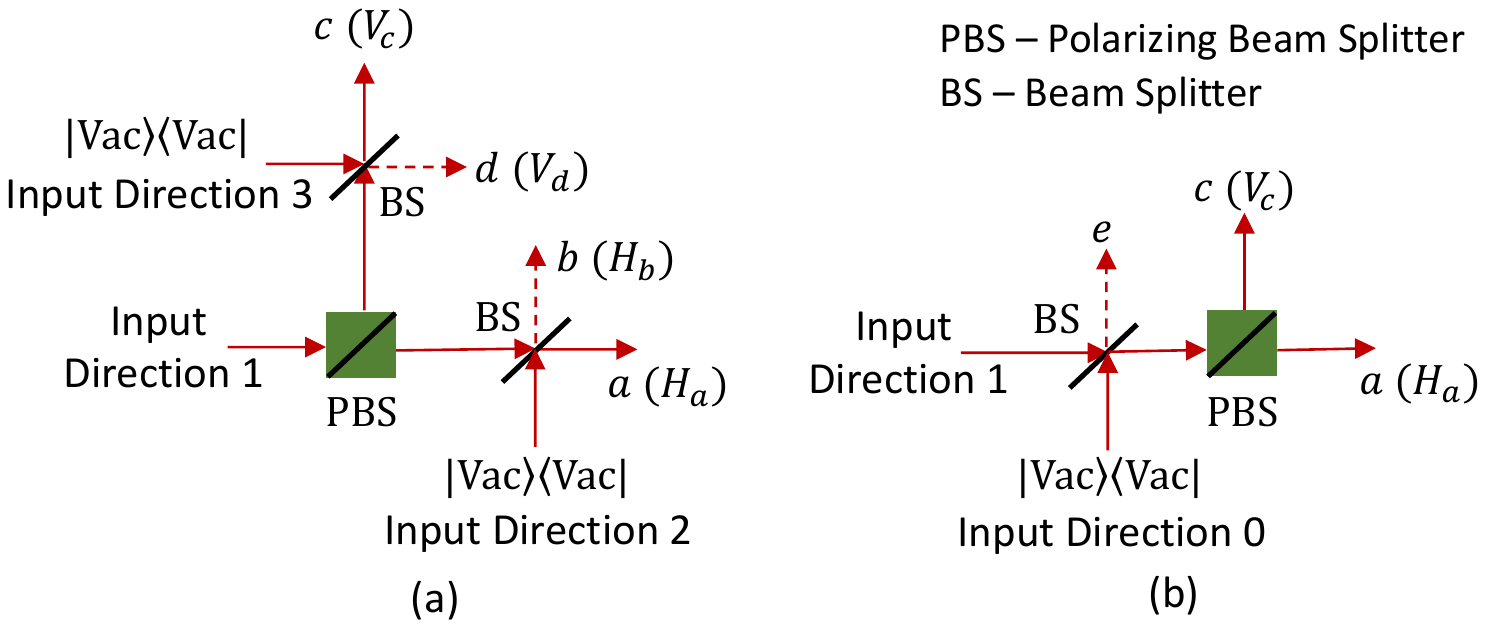} 
   \caption{Two equivalent descriptions of the measurement in the $H/V$ basis 
   under the active-detection scheme: (a) is the actual situation (the same as 
   Fig.~\ref{active_equivalent}(a)), and (b) is the hypothetical situation 
   (the same as Fig.~\ref{active_equivalent}(b)). Suppose that the detection 
   efficiencies associated with the outcomes $H$ and $V$ can be written as 
   $\eta_0\eta_1$ and $\eta_0\eta_2$ respectively, where $0\leq \eta_0, \eta_1, 
   \eta_2\leq 1$. Then, to realize the measurement, each beam splitter as shown 
   has a transmission coefficient $\sqrt{\eta_0}$, and the output modes $H_a$ and 
   $V_c$ are detected by real detectors with efficiencies $\eta_1$ and $\eta_2$, 
   respectively. The two real detectors are not shown for simplicity. Note that  
   the other output modes $H_b$ and $V_d$ in (a) and $H_e$ and $V_e$ in (b) are  
   not detected, corresponding to the loss in the measurement process.} 
   \label{active_equivalent2}
\end{figure}

Without loss of generality, we assume that the measurement is performed in the 
$H/V$ basis. From Appendix~\ref{sect:povm_with_mismatch}, we already know the POVM 
elements for the measurement described in Fig.~\ref{active_equivalent}(a),
i.e., Eqs.~\eqref{povm_active_h_input2} and~\eqref{eq:povm_active_hv_input}
with the replacement of $\eta_{H/D}$ and $\eta_{V/A}$ by $\eta_0\eta_1$ and 
$\eta_0\eta_2$, respectively. To prove the equivalence, we can derive the POVM 
elements for the measurement described in Fig.~\ref{active_equivalent}(b). However, 
this approach is quite lengthy. To simplify the proof and make the idea behind 
clear, we take another approach.
 
It is straightforward to see that Fig.~\ref{active_equivalent}(a) (or 
Fig.~\ref{active_equivalent}(b)) is the same as Fig.~\ref{active_equivalent2}(a)
(or Fig.~\ref{active_equivalent2}(b)) where only the input spatial direction 1 has 
incoming optical signals and only the output modes $H_a$ and $V_c$ are detected
(by real detectors with efficiencies $\eta_1$ and $\eta_2$ respectively). The 
optical devices in Fig.~\ref{active_equivalent2}(a) and Fig.~\ref{active_equivalent2}(b) 
are equivalent because of the well-known result that the common loss $\eta_0$ in the 
two output arms of a polarizing beam splitter can also be introduced by inserting a 
beam splitter with a transmission coefficient $\sqrt{\eta_0}$ for both polarization 
modes into the input arm of the polarizing beam splitter.

\subsection{Proof of the equivalence of the descriptions in Fig.~\ref{passive_equivalent}}
\label{sect:proof_passive_equiva}

\begin{figure}[htb!]
   \includegraphics[scale=0.58,viewport=7cm 13cm 13cm 20.5cm]{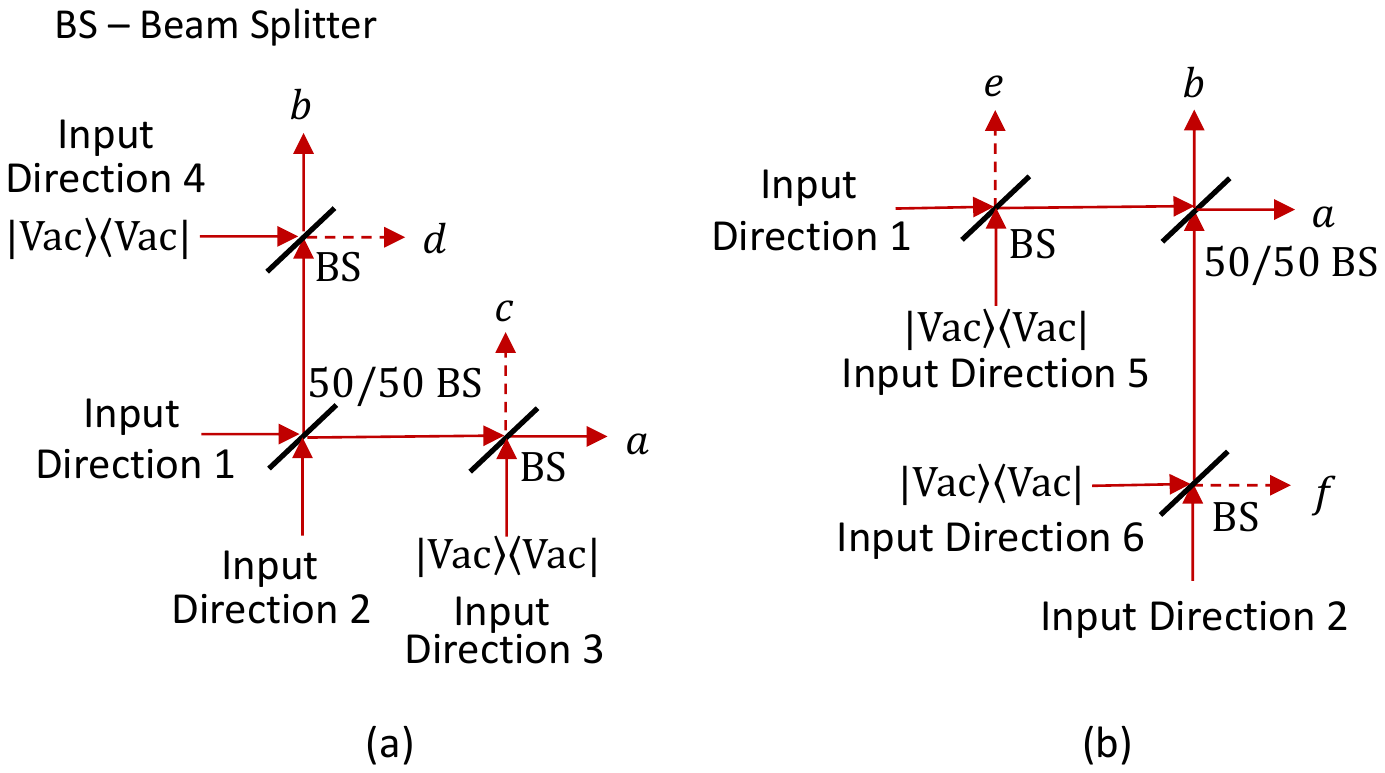} 
   \caption{Two equivalent descriptions of a $50/50$ beam splitter with a
   common loss in the two output arms: 
   (a) is the case where the loss is introduced in each output arm, and (b) is the 
   case where the loss is introduced in each input arm. Suppose that the common loss 
   is $\eta_0$. Then, each beam splitter as shown has a transmission 
   coefficient $\sqrt{\eta_0}$. The whole optical device in each case has four input 
   spatial directions and four output spatial directions.}
   \label{passive_equivalent2}
\end{figure}

In the passive-detection scheme as shown in Fig.~\ref{setup}(b), each output arm 
of the $50/50$ beam splitter is a setup for the active-detection scheme, i.e., a 
polarizing beam splitter with a detector in each output arm of the polarizing 
beam splitter. From Fig.~\ref{active_equivalent2} of Appendix~\ref{sect:proof_active_equiva}, 
we already see that a common loss in the two output arms of a polarizing beam splitter 
can be introduced either by inserting a beam splitter into each output arm or 
by inserting the same beam splitter into the input arm of the polarizing 
beam splitter. Hence, to prove the equivalence of Fig.~\ref{passive_equivalent}(a) and 
Fig.~\ref{passive_equivalent}(b), we only need to further prove that a common loss in 
the two output arms of a $50/50$ beam splitter as shown in Fig.~\ref{passive_equivalent2}(a) 
is equivalent to the same common loss in the two input arms of the $50/50$ beam splitter as 
shown in Fig.~\ref{passive_equivalent2}(b). This equivalence is well established in quantum 
optics, so here we skip the proof.

\subsection{Details of the SDP problem}
\label{sect:sdp_details}

In Sect.~\ref{sdp} of the main text, we formulate entanglement verification as 
a SDP feasibility problem of the form in Eq.~\eqref{eq:sdp}. Here we provide
more details of the SDP problem to be solved for the case of the active-detection 
scheme with one spatial mode. The SDP problems for the other cases studied in 
the paper can be formulated in the same way.

For the ideal BB84-QKD prepare-and-measure protocol studied in Sect.~\ref{toy_model}, 
Alice's system is equivalently described by a qubit, and her measurements are equivalently 
described by the $H/V$-basis measurement and the $D/A$-basis measurement (suppose that
information is encoded in the polarization degree of freedom). In the construction of 
an EVM $\chi$, to take advantage of the complete knowledge of Alice's state and  
measurements, we set the operators at Alice's side to be $\hat A_1=\Ket{\phi}\Bra{H}$ and 
$\hat A_2=\Ket{\phi}\Bra{V}$, where $\Ket{\phi}$ is an arbitrary qubit state. As discussed 
in Sects.~\ref{active_operators_1} and \ref{active_operators_2}, for Bob's side we consider 
both the actual measurement operators in Eq.~\eqref{eq:op_set} and the ideal operators 
in Eqs.~\eqref{eq:op_set0}, \eqref{eq:op_set1} and \eqref{eq:op_set2}. Denote the sets of 
operators exploited at Alice's side and Bob's side by $\mathbf{\mathcal{S}}^A$ and 
$\mathbf{\mathcal{S}}^B$, respectively. There are $2$ operators in the set $\mathbf{\mathcal{S}}^A$
and $18$ operators in the set $\mathbf{\mathcal{S}}^B$. Hence, the constructed EVM $\chi$ has
a dimension of $36\times 36$. We can constrain the EVM $\chi$ by the complete knowledge 
of Alice's state and measurements, by Alice's and Bob's observations $p_{AB}(x,y)$ with 
$x\in \{H,V,D,A\}$ and $y\in\{H,V,D,A, HV, DA\}$  (i.e., the probabilities that Alice observes 
outcomes $x$ while Bob observes outcomes $y$), by operator relationships, and by the bounds of 
the number of photons arriving at Bob as discussed in Sect.~\ref{photon_number_bounds}. In the 
following, we will discuss these constraints in detail.

First, let us discuss which entries of the EVM $\chi$ are fixed, given Alice's state 
$\rho_A$ and also the observations of Alice and Bob $p_{AB}(x,y)$. Recall that the EVM entries 
$\chi_{ij,kl}$ are the expectation values of the operators 
$\hat A_i^{\dagger}\hat A_k \otimes \hat B_j^{\dagger}\hat B_l$, where 
$\hat A_i, \hat A_k\in \mathbf{\mathcal{S}}^A$ and $\hat B_j, \hat B_l \in \mathbf{\mathcal{S}}^B$. 
So, we only need to figure out which operators' expectation values are known given $\rho_A$ and 
$p_{AB}(x,y)$. It is obvious to see that the expectation values of $\hat A_i^\dagger \hat A_k$ 
and $\hat A_i^\dagger \hat A_i \otimes \hat B_m$ are known, where $i,k=1,2$ and $\hat B_m$ 
denotes one of Bob's actual measurement operators in Eq.~\eqref{eq:op_set}. In total, there 
are 16 constraints of this type for the case of the active-detection scheme with one spatial mode.
Moreover, considering the relation 
$\hat A_1^\dagger \hat A_2=(\Ket{D}\Bra{D}-\Ket{A}\Bra{A}-\Ket{D}\Bra{A}+\Ket{A}\Bra{D})/2$ 
and the fact that both the actual measurement operators $\hat B_m$ of Bob and the joint state of Alice 
and Bob can be represented by real-valued matrices (see the discussion below Eq.~\eqref{rho_block}), 
it is easy to figure out the expectation values  of $\hat A_1^\dagger \hat A_2 \otimes \hat B_m$ 
given Alice's and Bob's observations. In the same way, we can also figure out the expectation   
values of $\hat A_2^\dagger \hat A_1 \otimes \hat B_m$. In total, there are 12 constraints 
of this type for the case of the active-detection scheme with one spatial mode.

Second, we discuss the equality constraints on the EVM entries $\chi_{ij,kl}$. There are 
three different kinds of equality constraints. The first kind is due to the relationships 
between operators exploited. For example, if two operators $\hat B_j$  and $\hat B_l$ commute and they 
are Hermitian, then by the definition in Eq.~\eqref{general_evm} the two EVM entries $\chi_{ij,kl}$ 
and $\chi_{il,kj}$ are equal to each other, $\forall i, k$. Particularly if the two operators 
$\hat B_j$ and $\hat B_l$ are orthogonal to each other (for example, $\hat B_j$ and $\hat B_l$ 
are operators in different photon-number subspaces), then the entries $\chi_{ij,kl}$ and 
$\chi_{il,kj}$ are zeros. We formulated 188 constraints of this type for the case of the 
active-detection scheme with one spatial mode. More equality constraints on the EVM entries 
can be derived by the commutation relationships as discussed in Appendix~\ref{sect:spin_operators}.  
We derived 72 constraints of this type for the case of the active-detection scheme with 
one spatial mode. The second kind of equality constraints is due to the fact that some operators 
exploited are represented by real-valued matrices. In this case, the corresponding EVM entries 
will be real numbers, since for entanglement verification we only need to consider the states 
$\rho_{AB}$ which are represented by real-valued matrices (see the discussion below Eq.~\eqref{rho_block}).
We formulated 34 constraints of this type for the case of the active-detection scheme 
with one spatial mode. The last kind of equality constraints is due to the fact that 
the projections of measurement POVM elements onto the $(\leq 2)$-photons subspace can be 
expressed as linear combinations of ideal operators in this subspace 
(for example, see Eq.~\eqref{eq:exp_upper_bound}). So, we can formulate the corresponding
equality constraints. For the case of the active-detection scheme with one spatial mode, 
there are 320 such constraints.

Third, let us consider the inequality constraints on the EVM entries $\chi_{ij,kl}$. 
First, such constraints can be derived from operator relationships. Suppose that 
$0\leq\hat B_j\hat B_l\leq \hat B_j$, $\hat B_j$ and $\hat B_l$ are Hermitian, and 
$\hat B_1$ is the identity operator in the full state space of Bob's system; then we 
have $0\leq \chi_{il,ij}, \chi_{ij,il}\leq \chi_{i1,ij}$ since the operators 
$\hat A_i^{\dagger}\hat A_i$ with $i=1,2$ are positive-semidefinite. These constraints can be 
derived from the operator relationships in Eq.~\eqref{eq:active_HV_rela}. In the same way, we 
can also derive inequality constraints from Eq.~\eqref{eq:exp_upper_bound2}. We formulated 
108 constraints of this type for the case of the active-detection scheme with one spatial mode.
Second, there is another way to derive inequality constraints, that is, by bounding the probabilities 
that the state lies in various photon-number subspaces. For example, we can derive an inequality 
constraint from Eq.~\eqref{eq:actual_doubleclick_bound}.  This is because given the observed 
double-click probability $d_{\text{obs}}$ and the optimized value $d_{3,min}$ all the other 
parameters $p_0, p_1, p_2$ and $p_2d_2$ in Eq.~\eqref{eq:actual_doubleclick_bound} can be 
written as linear combinations of EVM entries. In the same way, we can derive an inequality 
constraint from Eq.~\eqref{eq:actual_error_bound}. We formulated 16 constraints of this 
type for the case of the active-detection scheme with one spatial mode.

In a nutshell, by considering all the above linear constraints on the entries of the 
constructed EVM $\chi$, we can formulate entanglement verification as a SDP feasibility
problem of the form in Eq.~\eqref{eq:sdp}.

\bibliography{entanglement_verification_mismatch}

\end{document}